\documentclass{aa}

\usepackage{txfonts}
\usepackage{xcolor,varwidth}
\usepackage{lscape}
\usepackage{booktabs}
\usepackage{multirow}
\usepackage{amsmath}
\usepackage{longtable}
\usepackage{chemmacros}
\usepackage{graphicx}
\usepackage{siunitx}
\usepackage{multicol}
\usepackage{siunitx}
\usepackage{txfonts}
\usepackage{caption}
\usepackage{subcaption}
\usepackage[export]{adjustbox}
\newcommand{\cii}{[C\,{\sc ii}]}
\newcommand{\ci}{[C\,{\sc i}]}
\newcommand{\feii}{[Fe\,{\sc ii}]}
\newcommand{\neii}{[Ne\,{\sc ii}]}
\newcommand{\Neusi}{[S\,{\sc i}]}
\newcommand{\sii}{[S\,{\sc ii}]}
\newcommand{\hii}{\ion{H}{ii}}
\newcommand{\oi}{[O\,{\sc i}]}
\newcommand{\oiii}{[O\,{\sc iii}]}
\newcommand{\hi}{H\,{\sc i}}
\usepackage{hyperref}
\hypersetup{
  colorlinks = true, 
  urlcolor   = blue, 
  linkcolor  = blue, 
  citecolor  = blue 
}

\begin{document}

   \title{Dents in the Veil: Protostellar feedback in Orion}

   \author{\"U.~Kavak\inst{1,2,3,4}
          \and J.~Bally\inst{5}
          \and J.~R.~Goicoechea\inst{6}
          \and C.~H.~M.~Pabst\inst{4,6}
          \and F.~F.~S.~van~der~Tak\inst{2,1} 
          \and A.~G.~G.~M.~Tielens\inst{4}
          }
   \institute{Kapteyn Astronomical Institute, University of Groningen, P.O. Box 800, 9700 AV Groningen, The Netherlands \\
             \email{ukavak@sofia.usra.edu}
         \and
             SRON Netherlands Institute for Space Research, Landleven 12, 9747 AD Groningen, The Netherlands
         \and
             SOFIA Science Center, USRA, NASA Ames Research Center, M.S. N232-12, Moffett Field, CA 94035, USA
         \and
            Leiden Observatory, Leiden University, PO Box 9513, NL-2300RA, Leiden, the Netherlands
         \and
            Department of Astrophysical and Planetary Sciences, University of Colorado, Boulder, Colorado 80389, USA
         \and
             Instituto de Fisica Fundamental, CSIC, Calle Serrano 121-123, 28006 Madrid, Spain}

   \date{Received Month XX, 2021; accepted Month XX, 2021}

 
  \abstract
   {Interest in stellar feedback has recently increased because new studies suggest that radiative and mechanical feedback from young massive stars regulate the physical and chemical composition of the interstellar medium (ISM) significantly. Recent SOFIA \cii\,158~$\mu$m observations of the Orion Veil revealed that the expanding bubble is powered by stellar winds and influenced by previously active molecular outflows of ionizing massive stars.}
   {We aim to investigate the mechanical feedback on the whole Veil shell by searching for jets/outflows interacting with the Veil shell and determining the origin/driving mechanisms of these collisions.}
   {We make use of the \cii\,158~$\mu$m map of the Orion Nebula taken with the upGREAT instrument onboard SOFIA. We image the \cii\,emission the more extreme local standard of rest velocities ($v_\mathrm{LSR}$) between $-$3 and $-$20~km~s$^{-1}$ to pinpoint the high-velocity structures. Using position-velocity (PV) diagrams and high-velocity \cii\,emission, we search for spots of shock-accelerated \cii\,emitting gas, so called \textit{dents}. At these positions, we extract \cii\,line profiles to identify velocity components. We also compare the intensity distribution of the \cii\,emission with that of 8~$\mu$m PAH and 70~$\mu$m warm dust emission to see if there is a trend among these PDR tracers and to understand the origin of the dents.}
   {We identify six dents on the Veil shell with sizes between 0.3 and 1.35~pc and expansion velocities ranging from 4 to 14~km~s$^{-1}$ relative to the expanding Veil shell. The \cii\,line widths toward the dents vary from 4 to 16~km~s$^{-1}$ indicating that the dents are the result of interaction of highly turbulent motions (e.g., shocked gas) with the Veil shell. Moreover, dents appear only in the \cii\,PV diagram but not in the $^{12}$CO or \hi\,21~cm diagrams. Furthermore, the intensity distribution of the \cii\,emission of the dents has a tight correlation with that of 8~$\mu$m and 70~$\mu$m as long as the OMC or the Veil do not dominate its emission. Also, the observed dents do not have CO counterpart emission. These results indicate that the dents are made up of CO-dark H$_2$ gas. In the light of these findings, as well as the momenta of the dents and their dynamical timescales, we propose that the dents are created by the interaction of collimated jets/outflows from protostars with luminosities ranging from 10$^3$ to 10$^4$~$L_\odot$ indicating B-type stars in the Orion star-forming cloud with the surrounding Veil shell. However, it is challenging to pinpoint the driving stars as they may have moved from the original ejection points of the jets/outflows.}
   {We conclude that the dynamics of the expanding Veil shell is influenced not just by the O-type stars in the Trapezium cluster, but also by less massive stars, especially B-type, in the Orion Nebula. Mechanical feedback from protostars with a range of masses appears to play an important role in determining the morphology of \hii\,regions and injecting turbulence into the medium.}


   \maketitle
%

\section{Introduction}

    Interest in massive stars (with luminosities larger than 10$^3$~$L_\odot$, corresponding to a spectral type of B3 or earlier, and stellar masses higher than 8~$M_\odot$ \citep{Tan2014}) has increased in the last three decades as they inject considerable energy and momentum to unbind and disperse their natal clouds via stellar winds, powerful outflows, ionising radiation, and supernova explosions \citep{Krumholz2014, Bally2016, Motte2018}. The injection of mass, momentum, and energy which is called \textit{`stellar feedback'} can be seen on various spatial scales (from $\sim$1 to $\sim$100~pc) and dynamical timescales (from 10$^4$ to 10$^6$~years). At first glance, supernova explosions are the most energetic feedback process delivering immense energy (on the order of 10$^{51}$~erg seen in observations) that can reshape the morphology and composition of star-forming galaxies on large scales (10$-$100~pc) \citep{Thielemann2011}. However, recent studies reveal that feedback via protostellar outflows is also vital in setting the observed properties such as masses of stars \citep{Olivier2020, Guszejnov2021}.

    In contrast to their low-mass companions, massive stars reach their main-sequence luminosity while still embedded and accreting in a natal cloud of gas and dust due to their shorter Kelvin-Helmholtz timescales \citep{Zinnecker2007}. Protostars of all masses eject energetic jets/outflows to remove the angular momentum excess from the accretion process till they reach the main-sequence \citep{Beuther2002, SanchezMonge2013, Kavak2021}. This results in the entrainment of a significant amount of ambient molecular material. Even after the jets/outflows switch off when the star reaches the zero-age main sequence (ZAMS), relics of previously active molecular outflows, in other words fossil outflows, will continue to expand on their velocity vector and interact with the surrounding environment \citep{Quillen2005}. Furthermore, massive stars tend to form in dense clusters and exhibit a high multiplicity fraction \citep{Motte2018}. Therefore, it is possible to find newly forming massive stars and their outflows in the same cluster \citep{ODell2015} while other massive stars are already on the main sequence and radiate strong UV radiation \citep[as in the Orion Nebula; ][]{Bally2016}. From the observational point of view, quantifying the relative contribution of stellar feedback before and after reaching the ZAMS has been challenging for years \citep{Lopez2011}, despite the fact that state-of-the-art simulations are capable of employing stellar feedback modes individually \citep{Walch2012, Haid2018, Grudic2021}. 

    Orion's Veil (Veil for short), which is a series of foreground layers \cite[e.g., 9 layers identified by][]{Abel2019} of gas and dust lying in front of the Trapezium stars \citep{ODell2018}, is a unique laboratory for studying the relative effects of feedback mechanisms because its proximity allows us to resolve the bubbles in the Orion Molecular Cloud (OMC) spatially and and in Doppler velocity space \citep{ODell2011}. The \cii\,emitting Veil layer is a thin (0.5~pc) neutral and predominantly atomic shell \citep{Abel2016, Goicoechea2020} expanding at a velocity of 13~km~s$^{-1}$ toward us from the OMC-1 core driven by the kinetic energy converted from stellar winds of $\theta^1$~Ori~C, the most ionizing star in the Orion Nebula \citep{ODell2001, Pabst2019}. Some studies suggest a multi-layered structure model for the Veil based on the velocity components characterized through the emission and absorption lines \citep{Abel2016, ODell2018, Abel2019}. The main emission component of the Veil is traced by \cii\,fine-structure transition ($^2$P$_{3/2}$\,$\to$\,$^2$P$_{1/2}$ at 158~$\mu$m or 1.9~THz, i.e., $\Delta$E/$k_B$ = 91.2~K), which is the main cooling agent of neutral interstellar gas. While there are other tracers of CO-dark H$_2$ gas \citep[e.g., HF~$J$~=1-0;][]{Kavak2019}, \cii\,is by far the brightest as C$^+$ is the dominant carbon bearing species and the line is readily excited. Velocity-resolved \cii\,line observations are the state-of-the-art technique in determining the driving mechanisms of feedback in massive star-forming regions \citep{Goicoechea2015,YoungminSeo2019,Pabst2019, Pabst2020, Schneider2020, Tiwari2021, Luisi2021}. 

    Not only photoionization radiation from $\theta^1$~Ori~C, but high-velocity structures such as jets/outflows from YSOs and Herbig-Haro objects play a role in the dynamics of the Veil on various scales \citep{Henney2007, Bally2006, ODell1997}. Recently, \citet{Kavak2022} showed that even relics of previously active molecular outflows (i.e., fossil outflows) from $\theta^1$~Ori~C affect the morphology of the Veil. Blue-shifted ejections, which have relatively weak \oiii\,emission, are impinging on the neutral foreground Veil shell \citep[HH 202, HH 269, HH 203+204;][]{ODell2001} as the Veil itself expands. The collision of such objects with the Veil shell are a plausible explanation for the large temperature gradients \citep{Peimbert1991}. In this work, we investigate high velocity structures seen in the \cii\,observations using PV diagrams generated in cuts along the Veil and search for an association with shock accelerated atomic gas. Furthermore, we attempt to investigate the origin of the shock-accelerated gas\footnote{We use the term of \textit{dent} for the shock accelerated gas because the shocks collide with the inner surface of the Veil, resulting in hollow-like structures on the Veil's surface.} by estimating its momentum and dynamical timescale.

    We organize the paper as follows. In Section~\ref{Sect:Observations} we describe the \cii, $^{12}$CO and $^{13}$CO, and mid- and far-IR observations of the Veil. In Section~\ref{Sect:identificationdents}, we describe our methods to identify the dents on the Veil and to decompose the observed \cii\,line profiles over the dent position. Section~\ref{Sect:Analysis} contains an analysis of the momentum and energy of the dents. Finally, we discuss the origin of the dents and suggest possible further studies in Section~\ref{Sect:Discussion}.

\section{Observations}\label{Sect:Observations}

    \subsection{\cii\,observations}

    The observations were carried out with the Stratospheric Observatory for Infrared Astronomy (SOFIA), an airborne observatory project funded by the US National Aeronautics and Space Administration (NASA) and the German Aerospace Centre (DLR). SOFIA is a Boeing 747-SP jetliner that has been adapted to carry a 2.7-meter-diameter telescope in the back fuselage \citep{Young2012}.
    
    \begin{figure}
       \centering
       \includegraphics[width = \columnwidth]{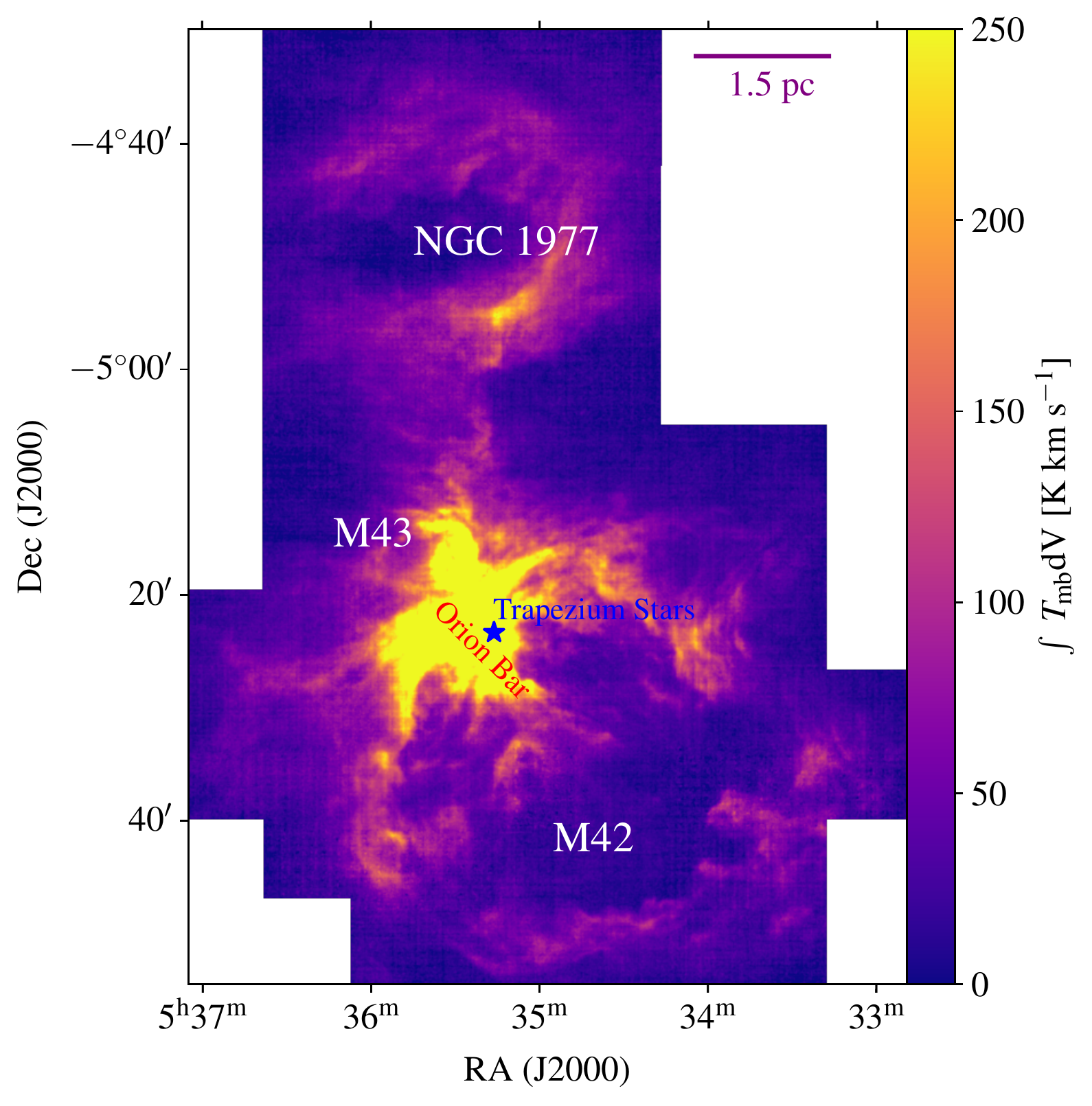}
       \caption{\cii\,158~$\mu$m integrated line intensity map (from $v_\mathrm{LSR}$ $-$50 and $+$50~km s$^{-1}$) of the OMC obtained with the upGREAT receiver onboard SOFIA. The bubbles of NGC~1977, M42, and M43 as well as the Trapezium stars and the Orion Bar are indicated. The 1.5~pc length is indicated by the line at the top-right.}
       \label{fig:cii_integratedmap}
   \end{figure}

    The data were collected with the German REceiver for Astronomy at Terahertz Frequencies (upGREAT) Instrument onboard SOFIA \citep{Risacher2018} for the Large program of the C$^+$ SQUAD led by A.~G.~G.~M.~Tielens. The spectral and spatial resolution during the observation is about 0.04~km~s$^{-1}$, and 14.1$\arcsec$. The final data were resampled to 0.3~km~s$^{-1}$ to achieve a better signal-to-noise ratio. The spatial resolution of the map is smoothed to $16\arcsec$, which corresponds to $\simeq$0.03~pc at the distance of Orion, 414~pc \citep{Menten2007}. The final rms noise (in $T_\mathrm{mb}$) is 1.14~K in 0.3~km~s$^{-1}$ velocity channels. More information on the observations and data reduction can be found in \citet{Pabst2020, Higgins2021}.
    
    \begin{figure*}
          \centering
          \includegraphics[width=\linewidth]{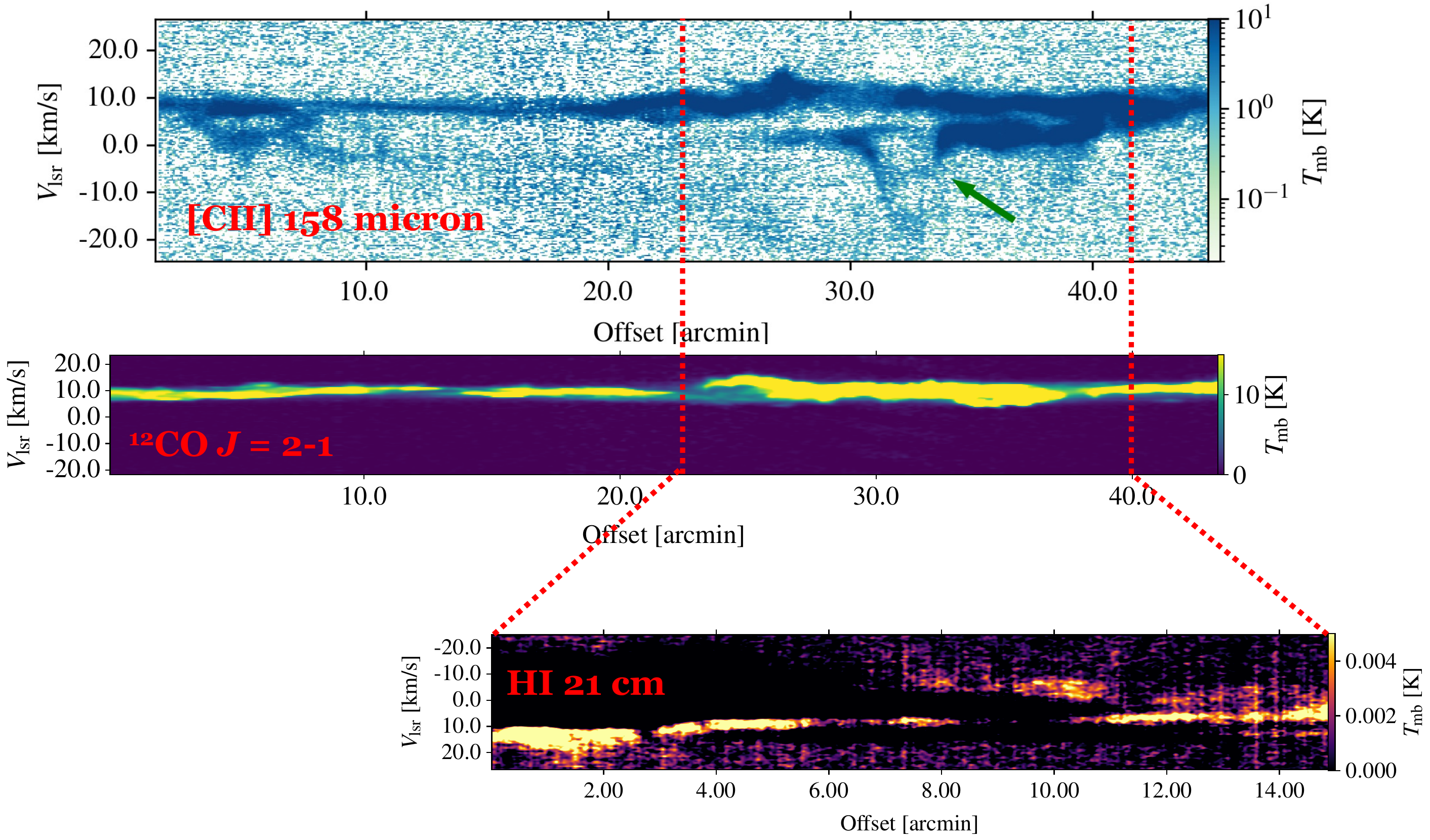}
        \caption{PV diagrams of dent 4. The emission at a $v_\mathrm{LSR}$ velocity of $+$9~km~s$^{-1}$ is the OMC, the weak \cii\,emission of Veil is at $v_\mathrm{LSR}$ = $-$4~km~s$^{-1}$. The dent at 33$\arcmin$ is indicated by a green arrow in the \cii\,158~$\mu$m PV-diagram. The $^{12}$CO and \hi\,PV diagrams are shown in the middle and lower panels, respectively. \hi\,observation from \citet{Vanderwerf2013} exist between the offset between two red-dashed lines, which 15$\arcmin$ long, in the \cii\,and CO observations. Dent 4 is located at 7.5$\arcmin$ in the \hi\,pv diagram.}
        \label{fig:dent4_4text}
    \end{figure*}
    
    \subsection{Molecular gas observations}
    
    We make use of $^{12}$CO \textit{J} = 2-1 (230.5~GHz) and $^{13}$CO \textit{J} = 2-1 (220.4~GHz) line maps taken with the IRAM~30m telescope in the framework of the Large Program `Dynamic and Radiative Feedback of Massive Stars' (PI: J.~R.~Goicoechea). In order to facilitate comparison with the velocity-resolved \cii\,map, we smoothed the $^{12}$CO (2-1) and $^{13}$CO (2-1) data to the angular resolution of the SOFIA \cii\,maps of $16\arcsec$. The average rms noise level in these maps is 0.16~K in 0.41~km~s$^{-1}$ velocity channels. A more detailed description of the CO observations can be found in \citet{Goicoechea2020}.
    
    \subsection{Atomic gas observations}
    
    We use \hi\,21~cm observation of the Veil shell obtained with the Karl G. Jansky Very Large Array at C and K configurations. The observation has an angular resolution of 7.2$\arcsec$~$\times$~5.7$\arcsec$ and a velocity resolution of 0.77~km~s$^{-1}$. Further details on data reduction and observation can be found in \cite{Vanderwerf2013}. 
    
    \subsection{Mid-IR observations}
    
    Mid$-$infrared observations were taken with the space-borne Spitzer telescope \citep{Werner2004} that conducted scientific observations between 2003 and 2020 with three focal plane instruments, one of which being the Infrared Array Camera \citep[IRAC;][]{Fazio2004}. IRAC is a four-channel camera that produces 5.2~$\times$~5.2 arcminute images at 3.6, 4.5, 5.8, and 8~$\mu$m. We utilize Spitzer 8~$\mu$m observations of the Orion Nebula to trace the UV-illuminated surface of the Veil. The FWHM of the point spread function is 1.9$\arcsec$ at 8.0~$\mu$m.

    \subsection{Far-IR photometric observations}
    
    The OMC has been observed as part of the Gould Belt Survey \citep{Andre2010} in parallel mode using the Photoconductor Array Camera and Spectrometer \citep[(PACS),][]{Poglitsch2010} and Spectral and Photo-metric Imaging Receiver \citep[(SPIRE),][]{Griffin2010} instruments on-board \textit{Herschel}. We use only the archival photometric images of PACS instrument at 70~$\mu$m tracing emission from warm dust grains for comparison with the \cii\,map over the dents.

\section{Identification of dents}\label{Sect:identificationdents}

\begin{figure}[!ht]
            \centering
            \includegraphics[width = \columnwidth]{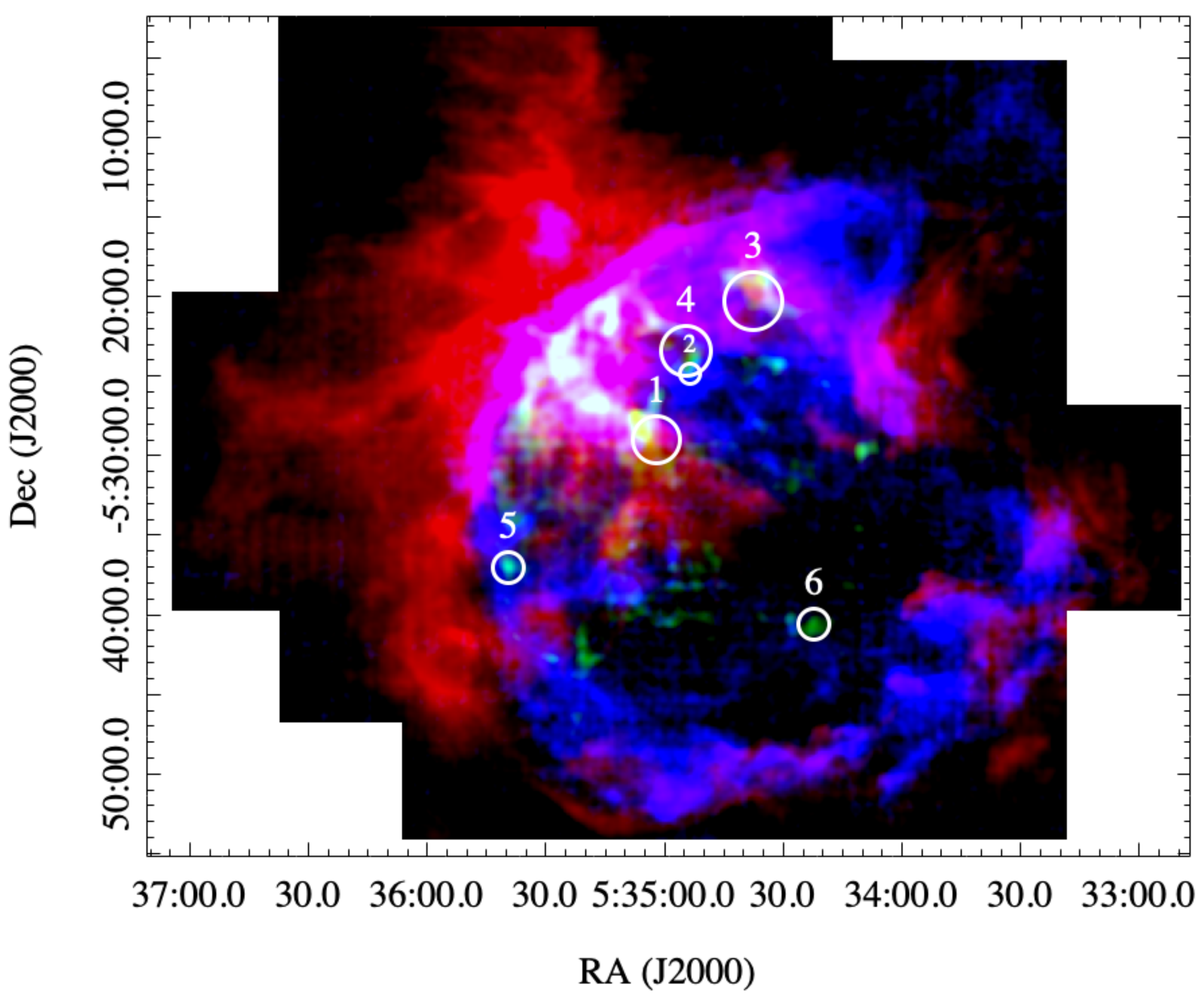}
            \caption{Three-color map of \cii\,emission in the Orion Nebula. Red hue represents the integrated \cii\,emission from the OMC between $+$20 and $+$3~km~s$^{-1}$. Blue represents the blue-shifted \cii\,emission generated by the Veil shell moving between $+$3 and $-$3~km~s$^{-1}$. Green represents high-velocity \cii\,emitting gas at velocities ranging from $-$3 to $-$20~km~s$^{-1}$. Gaussian smoothing of radius 35$\arcsec$ is performed to all three colors to reduce the image noise. The white circles indicate the position and size of the dents measured via PV diagrams that is also the aperture size for extracting \cii\,line profiles of the dents. More information is given in Section~\ref{Sect:Analysis}.}
            \label{fig:highvelocity}
        \end{figure}

        
    The dents can be detected in velocity-resolved channel maps and position-velocity diagrams \citep{Quillen2005}. We first identify notable dents in \cii\,PV diagrams along the Orion Veil (Section~\ref{sect:pvdiagrams}). We find that the dents emit at a $v_\mathrm{LSR}$ of $-$3 to $-$20~km~s$^{-1}$, which is about $-$15 to $-$30~km~s$^{-1}$ blue-shifted from the OMC emission (Section~\ref{sect:spectrumofdents}). We then integrate the \cii\,emission (green hue in Fig.~\ref{fig:highvelocity}) between these velocities to identify further dents in high-velocity \cii\,channels (Section~\ref{sect:highvelocity}).
        
\subsection{Position-velocity (PV) Diagrams}\label{sect:pvdiagrams}
        
    Because structures in the Veil are hard to find in the integrated map of \cii, the unbiased way of identifying shock-accelerated material or dents is the PV diagram. We examine \cii\,PV diagrams of the Orion Veil produced with 30$\arcsec$ broad horizontal and vertical slices. The horizontal cuts (east-to-west) are 60$\arcmin$ long, while the vertical cuts (south-to-north) are 45$\arcmin$ long because of the non-spherical morphology of the Orion Nebula.
        
    PV diagrams uncover the complicated structure of the Veil exposed to ionizing radiation from the Trapezium stars \citep{ODell2017}. In all PV diagrams, the \cii\,emission at $v_\mathrm{LSR}$ = $+$9~km~s$^{-1}$ indicates the background cloud OMC-1 (see PV diagram of dent 4 in Fig.~\ref{fig:dent4_4text} and of all dents in Appendix~\ref{sect:dentpvdiagrams}). The main blue-shifted expanding structure, moving towards us, is the Veil shell, expanding at 13~km~s$^{-1}$ \citep{Pabst2019}. In addition to these structures, we find `V-shaped' substructures that expand faster than the Veil. Using \cii\,PV diagrams, we identify four dents in the Veil, which are listed with their properties in Table~\ref{t:dents}. The expansion velocities are measured relative to the blue-shifted Veil shell. To this end, we extract the peak velocity, which is determined via Gaussian fitting of the dent spectrum, of the dent from that of the Veil component. The average size of the dents is about 0.3~pc, which is equal to 2.5$\arcmin$. The size of the dent is calculated along the RA axis in the PV diagrams. Since we know the width of each crosscut, the size of the dent in Dec is estimated by the number of PV diagrams in which it appears.
        
    Additionally, we examine the PV diagrams of the \hi\,21~cm and $^{12}$CO observations of the Veil shell to see if the dent looks identical to that of \cii\,158~$\mu$m (see also Fig.~\ref{fig:dent4_4text}). First of all, we find that \citet{Vanderwerf2013}'s \hi\,observation covers four of the dents identified before in the \cii\,PV diagrams. In Appendix~\ref{sect:dentpvdiagrams}, we provide \hi\,PV diagrams of these dents, respectively. Because the \hi\,PV diagrams behave differently, the \hi\,observations cannot be used to trace the dents.

        \begin{table*}[ht!]
            \small
            \centering
            \caption{Dents identified in this work. The location and sizes of the dents are shown in Fig.~\ref{fig:highvelocity}.}
            \label{t:dents}
            \begin{tabular}{c c c c c c c c c c c}
            \hline\hline
            
            & RA~(J2000)
            & Dec~(J2000)
            & $\int$~$T_\mathrm{mb}d$v
            & Size$^a$
            & \multicolumn{2}{c}{$I_\mathrm{[CII]}^b$}
            & (f, $\alpha$)$^c$
            & Mass
            & $V_\mathrm{exp, dent}$
            & P$_\mathrm{dent}$
            \\
            \cline{6-7}
            ID
            & (h:m:s)
            & (\degr:\arcmin:\arcsec)
            & [K~km~s$^{-1}$]
            & (pc~$\times$~pc)
            & ($\times$~Veil)
            & ($\times$~OMC)
            & $d_\mathrm{2.0~pc}$
            & [$M_\odot$]
            & [km~s$^{-1}$]
            & [$M_\odot$~km~s$^{-1}$]
            \\
            \hline\hline
            1  & $+$05:35:01.5 & $-$05:29:03.5 & $5.27\pm1.51$ & 0.36~$\times$~0.24 & 0.65 & 0.04 & 5, 10 & $3.3\pm0.6$ & $6.40\pm1.1$  & $21.5\pm1.3$ \\ 
            2  & $+$05:34:53.4 & $-$05:24:56.7 & $9.91\pm4.23$ & 0.16~$\times$~0.18 & 0.28 & 0.33 & 12, 5 & $0.6\pm0.1$ & $9.60\pm3.4$  & $6.00\pm0.9$ \\ 
            3  & $+$05:34:37.4 & $-$05:20:19.9 & $11.6\pm3.18$ & 0.43~$\times$~0.30 & 0.62 & 0.11 & 5, 12 & $4.8\pm0.9$ & $8.00\pm1.7$  & $38.4\pm3.2$ \\ 
            4  & $+$05:34:54.2 & $-$05:23:30.5 & $6.27\pm1.50$ & 0.39~$\times$~0.30 & 0.20 & 0.12 & 5, 11 & $3.9\pm0.8$ & $13.5\pm1.7$  & $52.5\pm3.1$\\ 
            5 & $+$05:35:39.2 & $-$05:37:05.2 & $22.8\pm1.25$ & 0.12~$\times$~0.12  & 1.35 & 1.00 & 16, 3 & $1.4\pm0.3$ & $9.20\pm0.6$  & >$13.6\pm0.7$ \\
            6 & $+$05:34:22.0 & $-$05:40:36.7 & $6.23\pm3.61$ & 0.24~$\times$~0.24  & 0.92 & 0.46 & 16, 3 & $1.5\pm0.3$ & $4.30\pm2.3$  & >$6.40\pm2.0$ \\
            \hline 
            \end{tabular}
            \tablefoot{
                    \tablefoottext{a}{One arcminute corresponds to the physical size of 0.12~pc at the distance of the Orion Nebula \citep[414~pc;][]{Menten2007}. The error in size is around 10\%.}
                    \tablefoottext{b}{$I_\mathrm{[CII]}$ denotes the integrated intensities of the dent. The values indicate which component dominates \cii\,emission at the dent position. If the value in the Veil and the OMC columns are >~1, that dent has brighter \cii\,emission. For fit results, see Table~\ref{t:fitresults}.}
                    \tablefoottext{c}{$d$, f, and $\alpha$ denote the distance between star and the Veil surface, the collimation factor and opening angle (in degree) of possible outflows, respectively. f and $\alpha$ are given for distances of 2.0~pc. See Section~\ref{Sect:Analysis} for more detail.}
                    }
        \end{table*}

\subsection{High-velocity \cii\,emission}\label{sect:highvelocity}
        
    As an alternative approach to identifying the interaction of stellar jets/outflows with the Veil, we peruse the \cii\,channel maps. Inspection of these maps reveals \cii\,emission between $-$20 and $-$3~km~s$^{-1}$. We examine the \cii\,channel maps, which show blue-shifted gas with a rather high velocity associated with the Veil. This is accomplished by superimposing the \cii\,emission within this velocity range as a green hue across the \cii\,of emission of the Veil (blue hue) and the OMC-1 background cloud (red hue) in Figure~\ref{fig:highvelocity}.

    This procedure also results in identifying dents 1$-$4. In addition, two more locations of high-velocity \cii\,emission become apparent (dents 5 and 6 in Fig.~\ref{fig:highvelocity}). These spots are the brightest and farthest high-velocity gases from the Trapezium cluster, respectively. The behaviour of these high-velocity structures does not appear as a dent in the PV diagrams (see Section~\ref{sect:pvdiagrams}). Examining the consecutive PV diagrams covering dents 5 and 6 shows that there is blue-shifted emission slightly faster than the Veil and that the velocity of the \cii\,emitting gas increases towards the peak of these structures (see Figs.~\ref{fig:Dent5} and \ref{fig:Dent6}).

    Note also that there is high-velocity \cii\,emission we have not considered in this work, especially in the Huygens region. This region has been the subject of many publications: \cite{Vanderwerf2013, ODell2001, ODell2009, Abel2019}. The origin of this emission could be a combination of the radiative and mechanical feedback from the Trapezium cluster and perhaps also from the stars in the Orion-S cloud \citep{ODell2009}. We also note that we limit this work focusing on dent 5 and 6 detected in high-velocity \cii\,emission, since other green hues, especially in Huygens region, exhibit very complex behavior and do not allow us to estimate the properties of the dents.
    
\subsection{Line profiles}\label{sect:spectrumofdents}
        
\begin{figure}[ht!]
            \centering
            \includegraphics[width = 0.85\columnwidth]{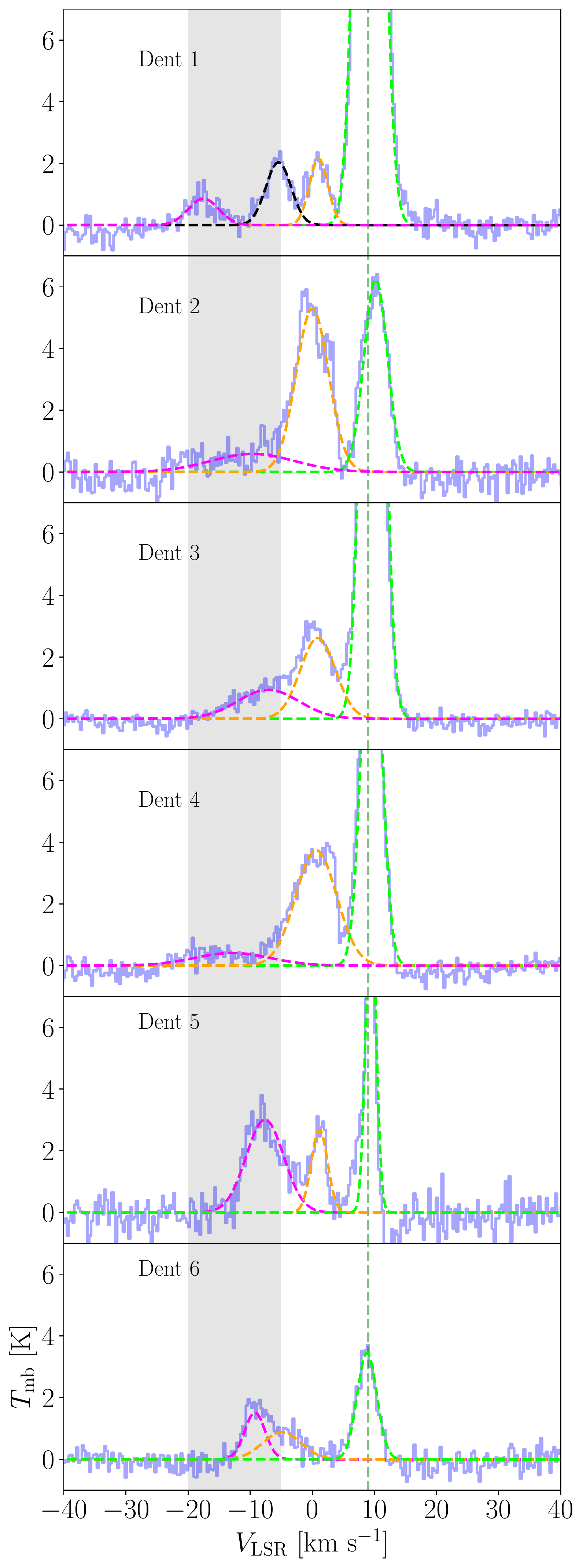}
            \caption{\cii\,158~$\mu$m line profiles using a circular extraction aperture of dent-size width towards the Veil dents listed in Table~\ref{t:dents}. The vertical green-dashed lines indicate the system velocity ($+$9~km~s$^{-1}$) of Orion. The light-green Gaussian fitting line is the OMC, the orange fitting line is the Veil, and the magenta fitting line is the dent. The black is component observed at the $v_\mathrm{LSR}$ of $-$5~km~s$^{-1}$ is a velocity component seen at the south of the Orion Bar PDR in the Huygens region.}
        \label{fig:dentspectrum}
        \end{figure}

    The \cii\,emission towards the dents shows a complex structure in the PV diagrams. To explore the origin of each component, we extract the \cii\,spectral line profiles across the six dents from the data cube and present them in Fig.~\ref{fig:dentspectrum} between LSR velocities of $-$30 and $+$30~km~s$^{-1}$. We utilize the size of the dent in arcminutes estimated from the PV diagrams to determine the size of the extraction region. In the direction of the dents, the line profile suggests a multi-component structure.
        
    We used a multi-Gaussian model to fit the \cii\,spectra and estimate the line parameters. The fit results are listed in Table~\ref{t:fitresults}. In the local standard of rest, all \cii\,spectra exhibit three major components: (i) the OMC at $+$9~km~s$^{-1}$, (ii) Orion's Veil at about $-$2~km~s$^{-1}$, and (iii) the dents at $-$10~km~s$^{-1}$. Dent 1 is an outlier, since it exhibits a double-peak at the Veil's velocity (i.e., black and orange fits) as well as a component at the extreme velocity of $-$19~km~s$^{-1}$ (see black Gaussian fit in Fig.~\ref{fig:dentspectrum}).


    Dents 5 and 6 behave slightly different than the other dents (Figs.~\ref{fig:Dent5} and \ref{fig:Dent6}). The PV diagrams of these dents appear to have a bright head of emission that is not linked with the Veil at first glance (see the PV diagram at the top in Figures~\ref{fig:Dent5} and \ref{fig:Dent6}). In addition, their brightness is similar to the Veil's. We check the spectra of the adjacent places to see if accelerated gas is present at the dent positions. According to the spectra in Fig.~\ref{fig:spectrum5_6}, these two dents are expanding somewhat faster than the Veil. In Section~\ref{Sect:Analysis}, we provide information regarding the origin of the \cii\,emission at the head of the dents.

\section{Analysis}\label{Sect:Analysis}

    In Section~\ref{sect:characteristics}, we summarize the properties of the dents. The momentum, which is the key parameter used to analyze the driving mechanism, of each dent is then estimated (Section~\ref{sect:momentum}) to comprehend the driving process. We compare the \cii\,emission with two crucial PDR tracers in Section~\ref{sect:originofdents}. Finally, we discuss a possible tracer of dent-like features on the ionization front of \hii\,regions in Section~\ref{sect:denttracer}.

\subsection{Characteristics of the dents}\label{sect:characteristics}

    In the previous section, we identify six dents that have diameters ranging from 0.16 to 0.43~pc. The first four dents in Table~\ref{t:dents} stand out in PV diagrams, but the last two dents require confirmation by high-velocity \cii\,emission maps. Four out of the six dents are detected near the Huygens region hosting the Trapezium cluster. The other two are in the direction of the extended Orion Nebula \citep[EON;][]{Gudel2008}. None of the dents appear in $^{12}$CO-PV diagrams (see Appendix~\ref{sect:dentpvdiagrams}) which suggest a low molecular gas fraction. Morphologically, all dents look like (shark) teeth in PV space and their $v_\mathrm{LSR}$ is more negative than that of the Veil shell. The expansion velocity of the dents relative to the Veil ranges from 4 to 14~km~s$^{-1}$. Based on this, we suggest that the dents are made up of CO-dark H$_2$ gas similar to what is seen in the Orion Bar \citep{vanderTak2012, Goicoechea2015, Kavak2019}. The formation of H$_2$ remains unclear, but it could be reformed in the shocked gas.

\subsection{Momentum of the dents}\label{sect:momentum}
        
    The momentum of the dent is estimated to determine the driving mechanism on the assumption that the momentum of outflows from protostars is conserved. In this regard, the mass and velocity of the dent must be estimated. We measure the expansion velocity relative to the Veil shell using the fit results of the \cii\,line profiles. To this end, we subtract the velocity of the dent from that of the Veil (see Table~\ref{t:fitresults}). The mass parameter is, however, rather uncertain because the Veil shell has density variations of up to a factor of ten and a low $N_\mathrm{H}$ of $\sim$10$^{21}$~cm$^{-2}$ \citep{Pabst2020} toward the line-of-sight. However, it is possible to make an estimation based on the mass calculation reported by \citet{Pabst2019}. 
        
    The mass accelerated by shocks from the Veil shell outward equals at least the mass entrained in each dent. As the size (2.7~pc) and gas mass (1500~$M_\odot$) of the Veil are known \citep{Pabst2020}, we can roughly calculate the surface mass density of the Veil to calculate the mass parameter assuming a half-sphere geometry with radius of 2.7~pc for the Veil. The surface density of the Veil is $\sim$30~$M_\odot$~pc$^{-2}$. We multiply the area of the dent with the surface density to estimate the shock-accelerated mass, in other words, the mass in the dents. The mass estimation and momentum of the dents are given in Table~\ref{t:dents}. We also note that Dents 5 and 6 are oblique to the surface of the Veil as shown by a series of PV-diagrams (see Figs.~\ref{fig:Dent5} and \ref{fig:Dent6}) in Appendix~\ref{sect:dentpvdiagrams}. We, therefore, give a lower limit for the momentum of these two dents in Table~\ref{t:dents}. 
        
        \begin{figure*}[!t]
            \centering
            \includegraphics[width = 1.8 \columnwidth]{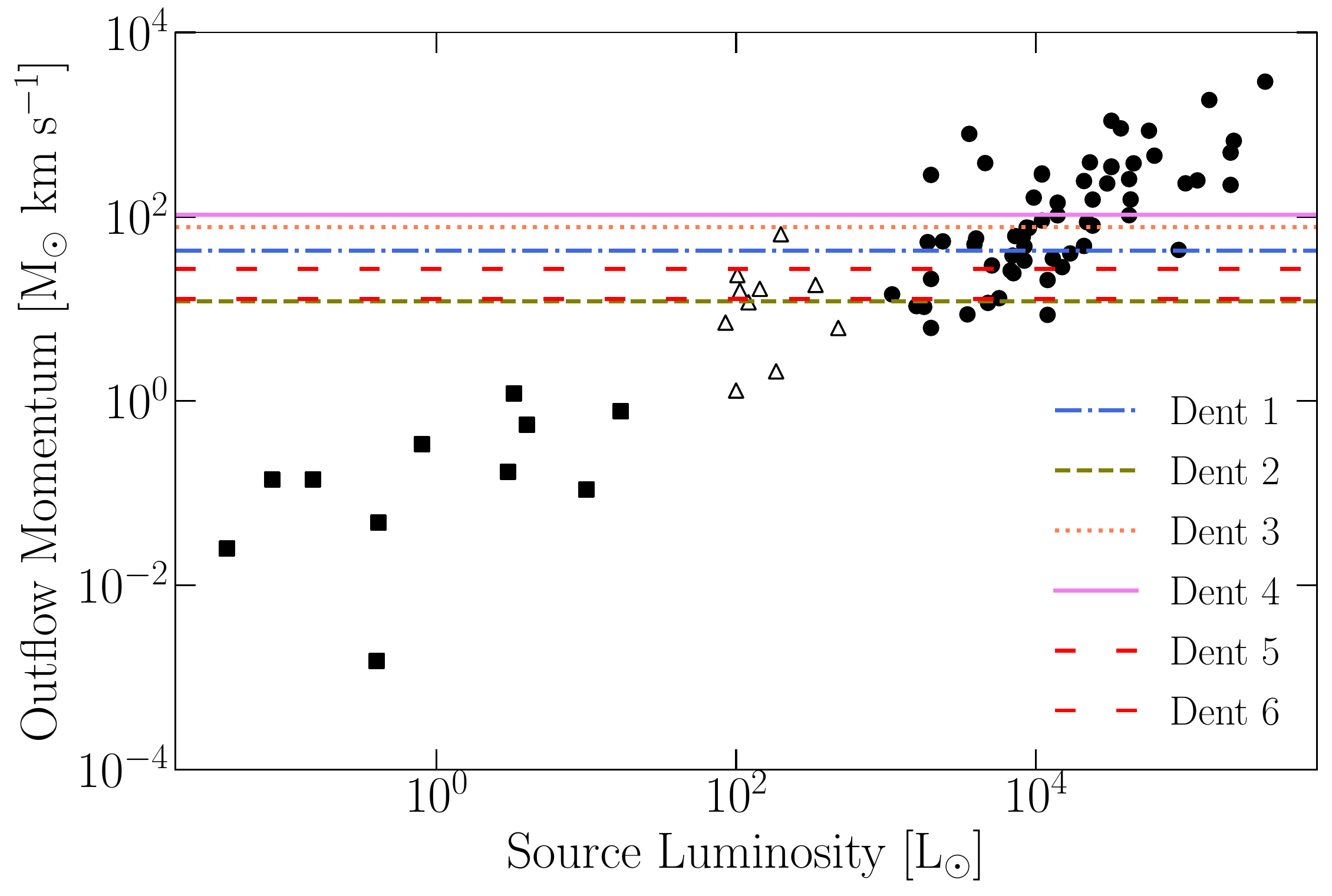}
            \caption{Plot of outflow momentum in $M_\odot$~km~s$^{-1}$ against the luminosity of its driving source in $L_\odot$. Black circles are from \citet{Maud2015}, and open triangles from \citet{DuarteCabral2013} for high-mass Class$-$0 objects, and black squares from \citet{Dunham2014} showing the outflows from low-mass stars. Horizontal lines denote the momentum (multiplied by two as outflows are bipolar) of each dent listed in Table~\ref{t:dents}. Dent 5 and 6 have a lower limit, and thus are indicated by red loosely-dashed lines.}
            \label{fig:OutflowMomentum}
        \end{figure*}
        
    The total momentum of the Veil is 18,000~$M_\odot$~km~s$^{-1}$ and the dents carry thus between 0.5 and 1\% of the total momentum injected by the Trapezium stars \citep{Pabst2020}. For comparison, the momentum contained in the protrusion to the northwest is 540~$M_\odot$~km~s$^{-1}$ that is 3\% of the momentum of the Veil shell \citep{Kavak2022}. 
        
    The correlation between the jet/outflow momentum and the luminosity of the protostars ejecting the material has been well established \citep{Bontemps1996, Wu2004, LopezSepulcre2010, DuarteCabral2013, SanchezMonge2013, SanJoseGarcia2013, Maud2015, Kavak2021}. These results indicate a relationship across the low- and high-mass regimes between these two quantities (see Fig.~\ref{fig:OutflowMomentum}). In this plot, each dent is individually marked by its momentum. The momentum of the dent implies massive stars of B-type with luminosities ranging between 10$^3$ and 10$^4$~$L_\odot$. Also, because protostellar jets and outflows are typically double lobed, we double the momentum predicted for each dent in our sample before comparing it to the protostellar activity correlation.
        
    Taking the size and the velocity as a guide, we estimate that the formation of the dents would take between (0.5$-$2.5)~$\times$~10$^{4}$~years. This represents $\sim$1/4 of the expansion timescale of the Veil shell \citep[$\sim$2~$\times$~10$^5$~years;][]{Pabst2019}, suggesting that the dents were produced during the expansion of the Veil by forming massive B- and A-type stars which is consistent with accreting massive stars reported by \citet{DuarteCabral2013}. As this young age indicates recent outflow activity, we consider jets/outflows from accreting massive protostars as the most likely driving mechanism.
        
\subsection{Origin of the dents}\label{sect:originofdents}
    
\begin{figure*}
            \centering 
            \includegraphics[width = 0.65\columnwidth]{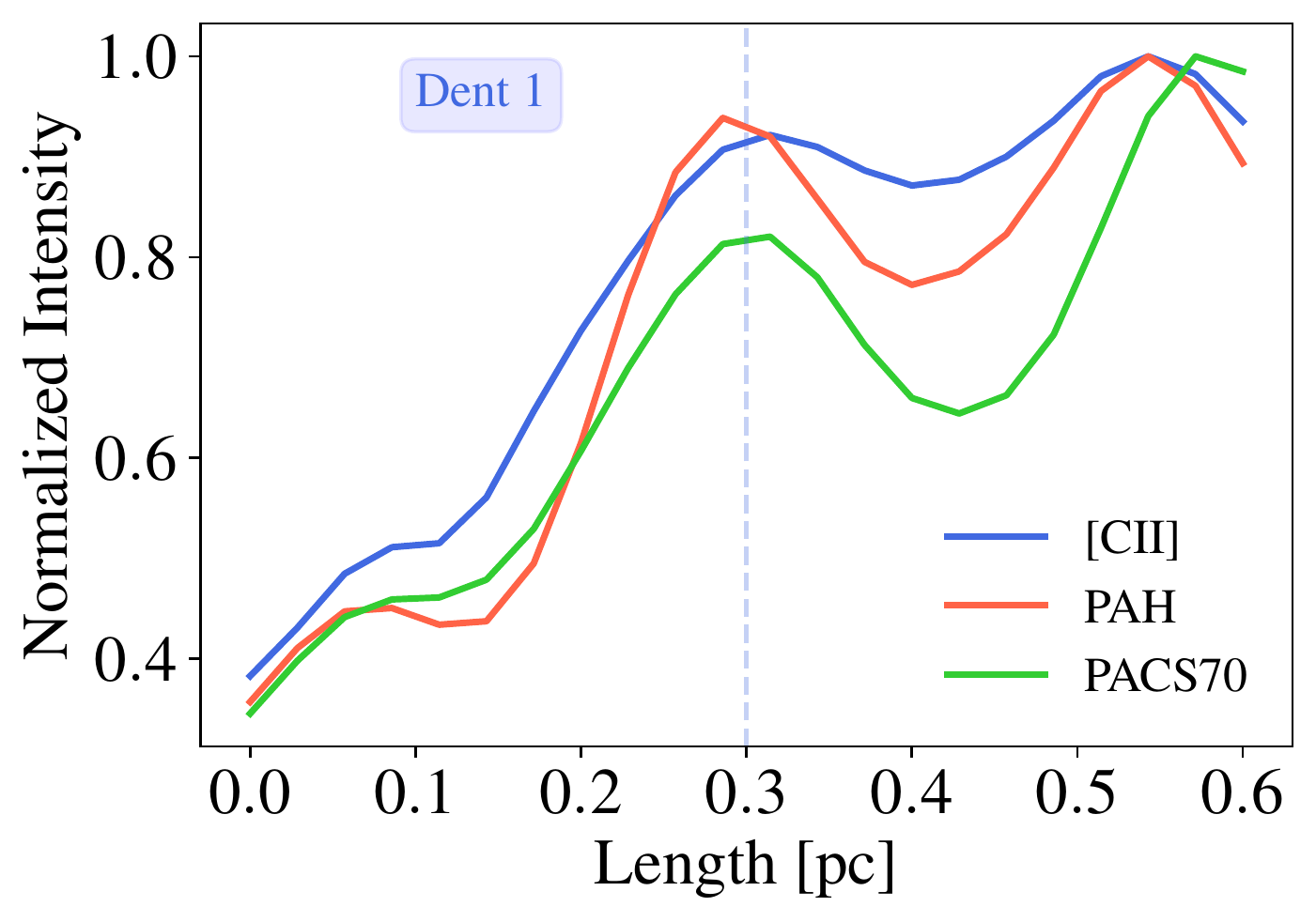}
            \includegraphics[width = 0.65\columnwidth]{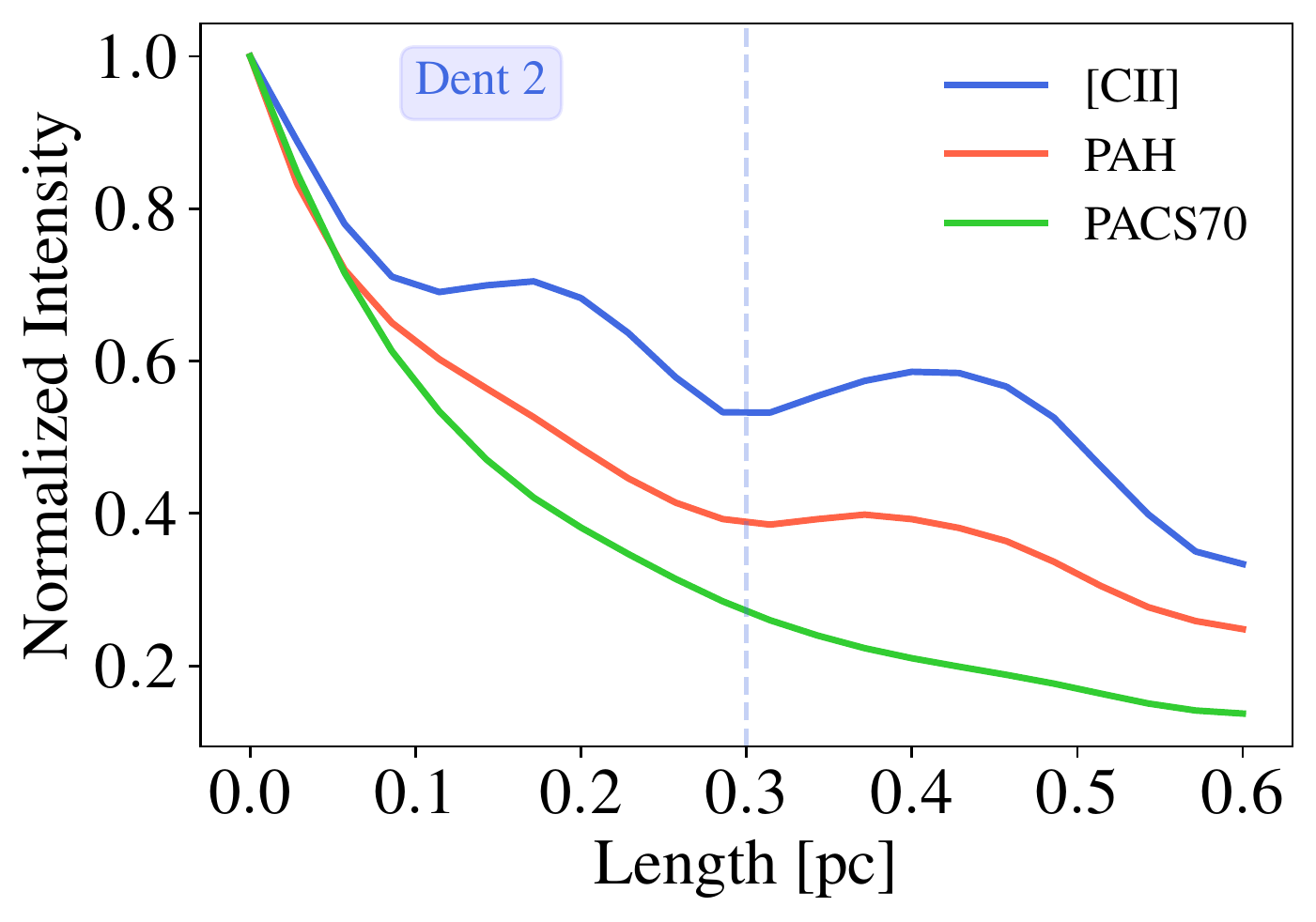}
            \includegraphics[width = 0.65\columnwidth]{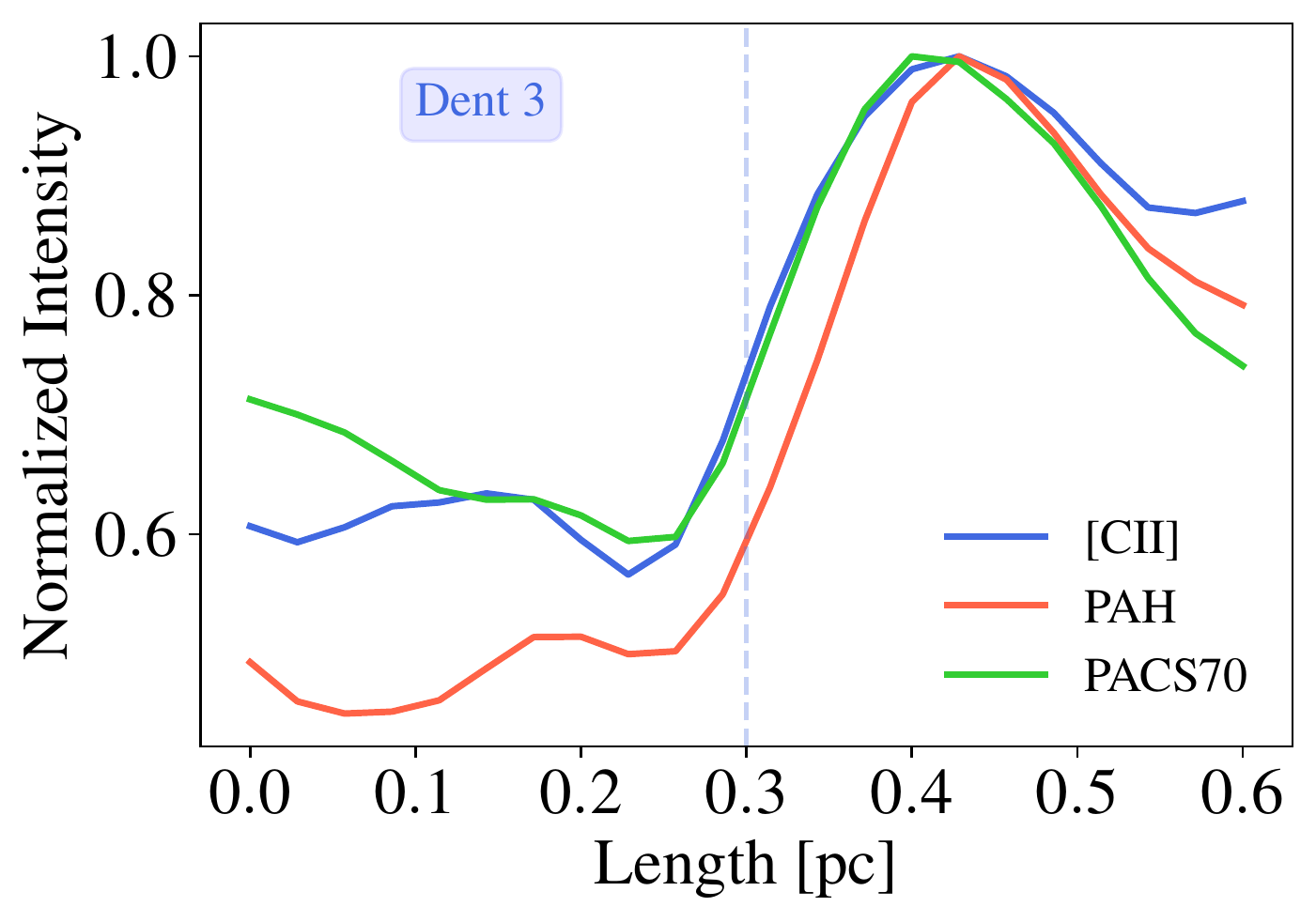}
            \includegraphics[width = 0.65\columnwidth]{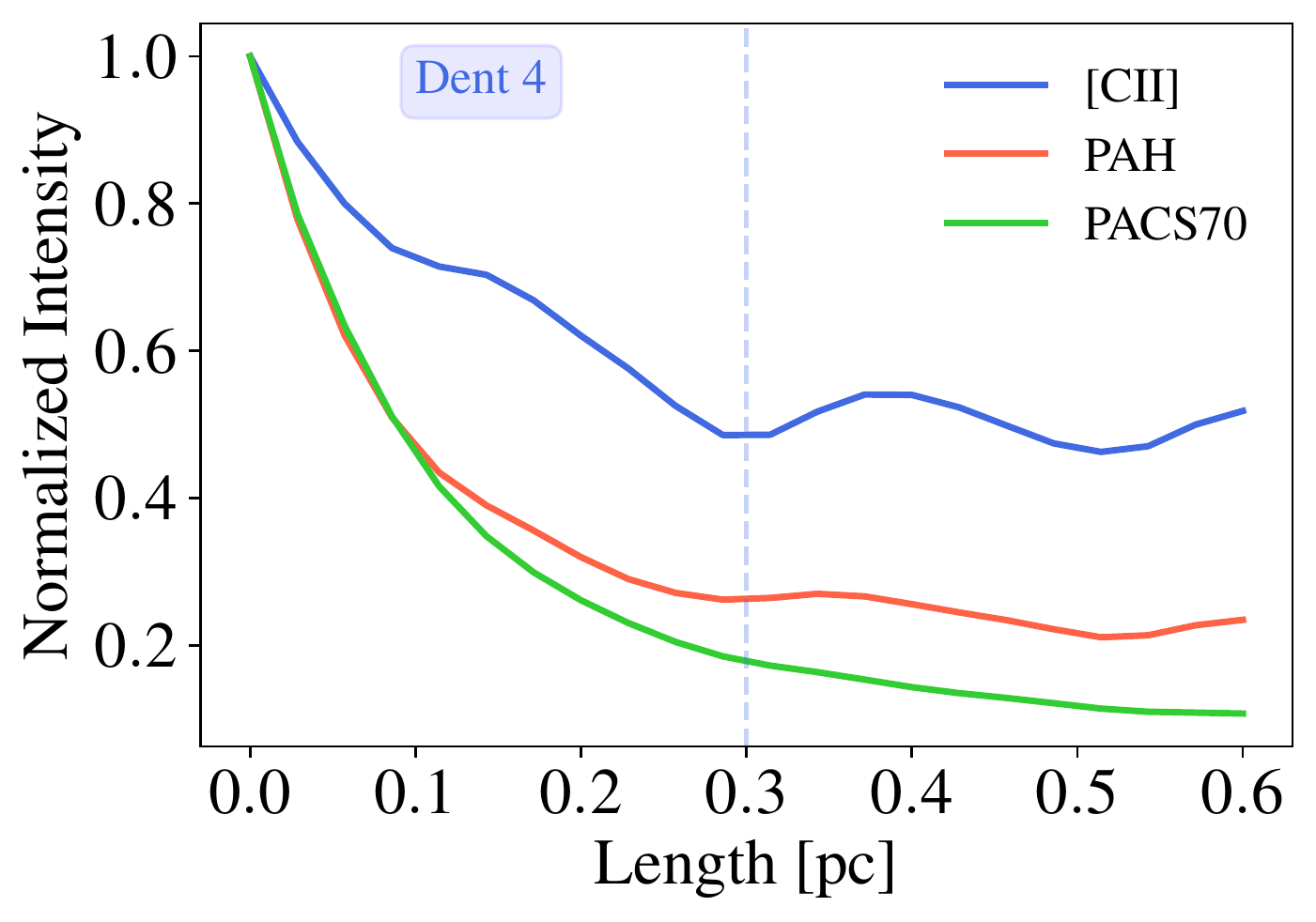}
            \includegraphics[width = 0.65\columnwidth]{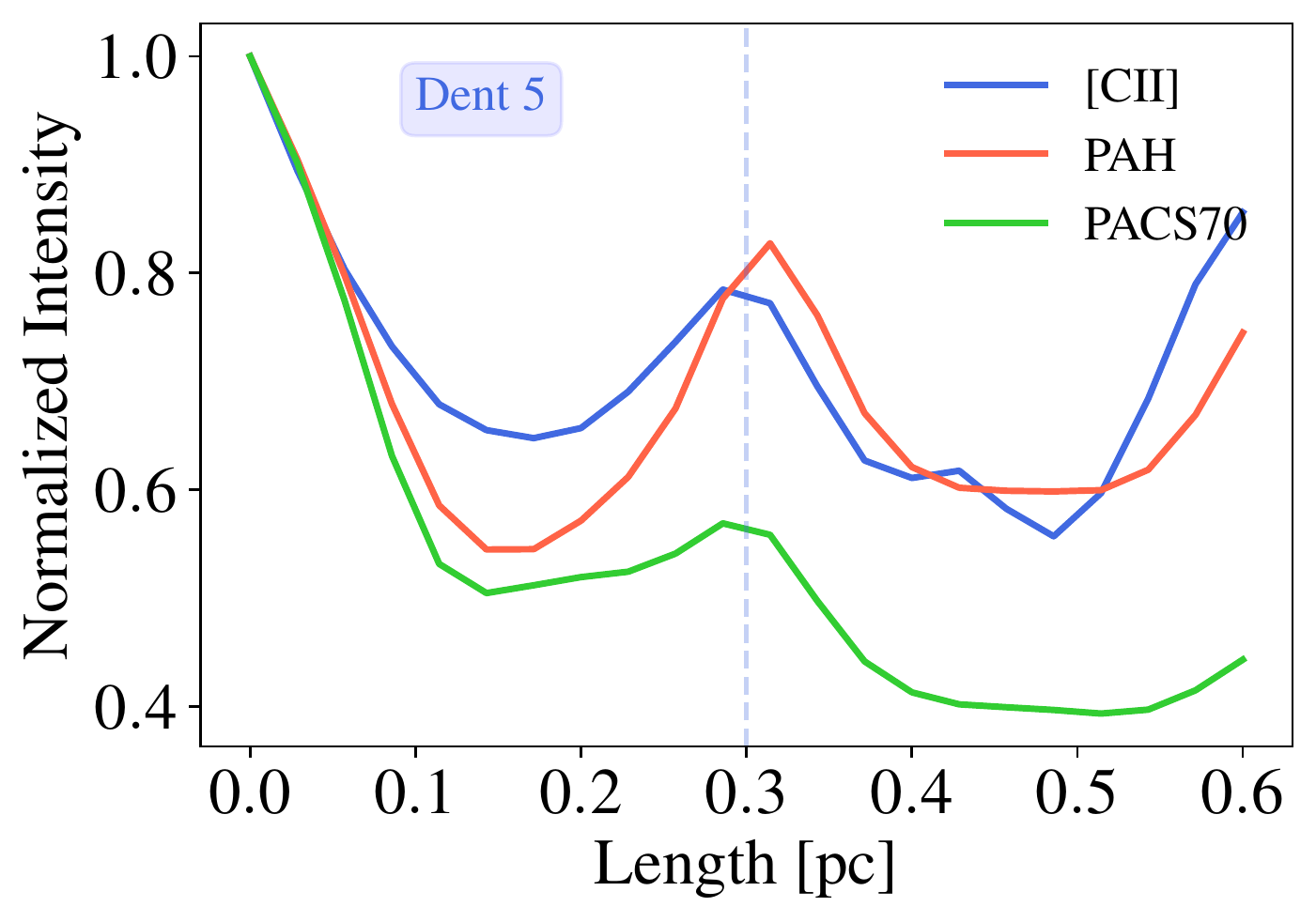}
            \includegraphics[width = 0.65\columnwidth]{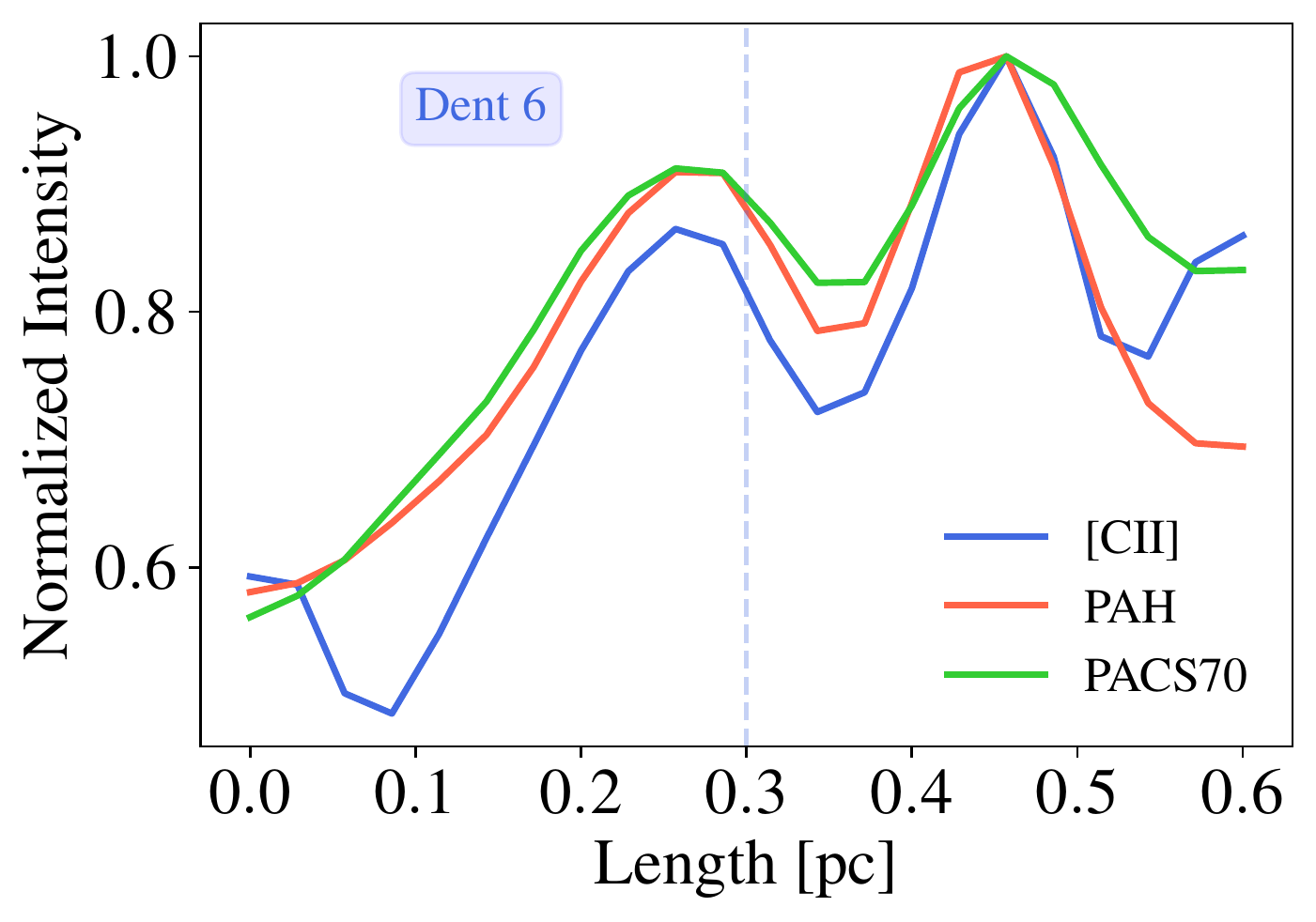}
            \caption{Comparison of normalized \cii\,intensity with to that of 8~$\mu$m and PACS~70~$\mu$m along 0.6~pc long horizontal cuts spanning the dents from east-to-west. The position of the dent is shown by the vertical blue-dashed line. All observations are convolved to $20\arcsec$ for a proper comparison.} 
            \label{fig:intensitycomparision}
        \end{figure*}

    We look for correlations between the dent \cii\,emission and tracers of UV versus shock illumination. For this, we make use of \cii\,emission with PDR tracers such as Spitzer 8~$\mu$m PAH emission and PACS 70~$\mu$m tracing warm dust (see Fig.~\ref{fig:intensitycomparision}).
        
    Dent 1 is located just south-west the bright Orion Bar PDR. The spectrum in the direction of the dent has four components (see Fig.~\ref{fig:dentspectrum}). The fitted lines in magenta, orange, and green represent the dent, the Veil shell, and background cloud OMC, respectively. The black component is observed at a $v_\mathrm{LSR}$ of $-$5~km~s$^{-1}$. This component has been interpreted as a broad arc in the shape of an incomplete semicircle near the border of the Huygens region, but the dent at $-$18~km~s$^{-1}$ is not seen in \hi\,21~cm observations \citep{Vanderwerf2013}. The semicircle structure host a group of stars including a B-star (see Fig.~\ref{fig:OBAstarsDents}). We suggest that dent 1 is accelerated by the jet/outflow of these stars to an $v_\mathrm{LSR}$ velocity of $+$20~km~s$^{-1}$. We also suggest that these dents reflect the interaction of jets/outflows with the expanding Veil nebula. The velocity of these jets/outflows must be in excess of the expansion velocity \cite[13~km/s;][]{Pabst2019} of the Veil itself. We will come back to this hypothesis in Section~\ref{Sect:Discussion}. 
    
    \citet{Luhman1994} concluded that  UV florescence dominates the extended IR H$_2$ (v>0) emission toward OMC. Indeed, the extended $K_s$ emission  (including the H$_2$ v=1-0 S(1) and v=2-1 S(1) lines) spatially correlates with the CH$^+$~$J$ = 1-0 emission \citep{Goicoechea2019}, a natural product of reactions between C$^+$ and UV-pumped H$_2$ (v>0) \citep[e.g.,][]{Agundez2010}. The extended Orion Nebula has been surveyed in the 1-0 S(1) line at 2.12~$\mu$m by \citep{Stanke2002}. We find that 12 out of 78 H$_2$ features are situated in the direction of the Orion Nebula but are unconnected to the dents because the structures reported by \citet{Stanke2002} are found in dense molecular clouds while dents are associated with the diffuse emission in the Orion Veil. We conclude that shocks revealing the interaction of protostellar jets with the Veil nebula are difficult to trace directly. The best signature seems to be the velocity shift and broader line-widths induced by this shock interaction in the atomic fine-structure cooling lines but, as argued above, the \cii\,emission is dominated by UV heated gas (see Section~\ref{sect:denttracer}).
        
\subsection{Potential shock signature of the dents}\label{sect:denttracer}

    If the dents are indeed due to a jet interacting with the Veil, then we would expect to see this region light up in typical shock tracers. Low velocity interstellar shocks can be J-type or C-shocks, depending on the strength of the magnetic field and the shock velocity \citep{Draine1983}. The line-of-sight magnetic field is measured to be $\sim$100 $\mu$G \citep{Troland2016}. For atomic gas, the critical velocity at which a C-shock becomes J-type is $\sim$20~km~s$^{-1}$ \citep{Lesaffre2013}. The observed velocities are consistent with a C-type shock in the range of 5 to 15~km~s$^{-1}$. Such a shock would heat a column density of $\sim$10$^{20}$ cm$^{-2}$ to $\sim$1000 K \citep{Lesaffre2013}. For a velocity in excess of ~20~km~s$^{-1}$, the shock would be J-type. The gas is then heated to $\sim$10$^5$~K in the shock front and, in the frame of the shock, would flow at 1/4 of the shock velocity. As the gas cools down, its velocity would decrease. In a J-type shock, cooling through atomic lines becomes more important.
        
    Comparing PDR models \citep{Kaufman2006, Pound2008} and shock models, for the atomic cooling lines (\cii, \oi, \ci, and etc.), the shock signature would be overwhelmed by the emission generated by the UV irradiation. The best tracers are low-$J$ H$_2$ lines \citep{Lesaffre2013} but there too, the UV$-$heated gas would have to be accounted for. As an example, the H$_2$ 0-0 S(1) intensity from a PDR with $G_0$~=~10$^2$ is predicted to be 10$^{-5}$ erg~cm$^{-2}$~s$^{-1}$~sr$^{-1}$; very comparable to the emission from a 10 km~s$^{-1}$ C-type shock and about 10 times the emission from a J-type shock \citep{Lesaffre2013}. High velocity J-type shocks will lead to emission in optical transitions such as \sii~$\lambda$6731. High velocity resolution will be required to separate this shock emission from photo-ionized gas in the extended Orion Nebula. Finally, we note that the substantial column density of warm gas in both C- and J-type shocks would enable reactions with substantial energy barriers to proceed and this could lead to detectable amounts of, for example, OH, and SH$^+$ \citep{Lesaffre2013, Godard2019}. As shocks heat the gas to much higher temperatures than PDRs, these species could be used as the signature of the presence of a shock. Similarly, high-spectral resolution observations of near-IR \feii\, and IR \Neusi\,and \neii\,lines could be used as shock tracers because these lines originate from levels that cannot be excited in low density, low UV field PDRs.
        
    In the dents, the \cii\,158~$\mu$m line-widths are always broader than those from the background OMC-1 cloud (see Table~\ref{t:fitresults}). Also, with the exception of Dents 5 and 6, the line-widths of the dent components are broader than that of the Veil, although the fit uncertainty is sometimes large. There could be a few possibilities to explain the broad line-widths: (a) the gas in the dents is warmer than the gas in the Veil, (b) or it is much more turbulent, as a result of the passage of the shock, (c) lines are broad because the \cii\,emission is from hot photoionised gas rather than neutral gas. The observed line width ($\sigma$) in the dents is,
        \begin{equation}
            \sigma = (\sigma_\mathrm{th}^2 + \sigma_\mathrm{turb}^2)^{1/2}
        \end{equation}
    where $\sigma_\mathrm{turb}$ is the non-thermal velocity dispersion and $\sigma_\mathrm{th}$ is thermal broadening which change as $(kT_k/m_{C+})^{1/2}$. This analysis is applicable to optically thin \cii\,emission, i.e., with no opacity broadening. First, we assume that $\sigma_\mathrm{turb}$ = 0. For the observed line-widths (with median value of 9.36~km~s$^{-1}$ in Table~\ref{t:fitresults}), we require a maximum temperature ($T_\mathrm{k}$) of the gas of 5~$\times$~10$^4$~K, which is significantly higher than than the neutral gas in Orion molecular cloud \citep{Goicoechea2015}. This may imply that \cii\,emission arises from a more turbulent gas in the Veil shell. To compute $\sigma_\mathrm{turb}$, we also assume that the \cii\,gas in the dents is at $T_\mathrm{k}$~$\sim$100~K \citep{Pabst2020}. At T$_k$~$\sim$100~K, the speed of sound is $\sim$1~km~s$^{-1}$, and $\sigma_\mathrm{turb}$ is 9~km~s$^{-1}$ for the line width of 9.36~km~s$^{-1}$ given in Table~\ref{t:fitresults}. This gives us non-thermal velocity distribution ($\sigma_\mathrm{turb}$), which is $\Delta$v$_\mathrm{FWHM}$/2.355, of $\sim$4~km~s$^{-1}$. This is greater than the speed of sound at 100~K indicating that nonthermal velocity distribution in the dents is as a result of turbulent motions such as shocked gas within jets/outflows.
\subsection{Collimation factor and opening angle}
    
    Assuming that the dents are driven by the jets/outflows of protostars, the collimation factor ($f$) may also be an indication of the type of star given that outflow collimation decreases from low to massive stars \citep{Bachiller1999}. However, outflows from B- or O-type stars can be well-collimated with factors higher than five on a dynamical timescale shorter than 10$^4$~yr \citep{Arce2007}. Moreover, \citet{Wu2004} report that the collimation factor of the outflow from a protostar with bolometric luminosity higher than 10$^3$~$L_\odot$ is about two. In our case, the degree of collimation can be estimated depending on the distance ($d$) between the star and the surface of the Veil shell (see also Fig.~\ref{fig:OBAstarsDents} for the assumed geometry). For this purpose, we assume that the star, which is powering the outflow is located in the core of the Orion Nebula cluster at a distance of 2~pc. The collimation factor varies between 5 and 12 while the opening angle ($\alpha$) is between 3 and 12$^{\degr}$ (see Table~\ref{t:dents}), indicating collimated ejections such as molecular jets from massive stars \citep{Arce2007}. If the star-dent distance were substantially smaller than the adopted value, the collimation factor would decrease and the opening angle would increase. For a distance of 0.5~pc, the typical collimation factor and opening angle would be 2 and 40$^\circ$, respectively. We estimate a timescale of 5.5~$\times$~10$^4$~years (see timescale above) for dent 1 involved in its formation, which is 1/4 of the expansion timescale of the Veil shell \citep{Pabst2019}. With this kind of a timescale, the star does not need to be directly behind the dent because a massive star with a proper motion of 2~km~s$^{-1}$ can travel about 0.1~pc ($\sim$0.8$\arcmin$) away in 5.5~$\times$~10$^4$~years from where its outflows were ejected. Therefore, estimating the driving stars of the dents is challenging.
        
    In addition, the \cii\,spectra around dents 5 and 6 suggest \cii\,emission from accelerated gas (see Fig.~\ref{fig:spectrum5_6}). Only these two dents show an increase in \cii\,brightness at the head of the dents. By comparing the intensities in Fig.~\ref{fig:intensitycomparision} and previous findings in Section~\ref{Sect:identificationdents}, we argue that these two dents are also formed by the same mechanisms on the surface of the Veil shell. We, however, are unable to find the same association for all dents in Fig.~\ref{fig:intensitycomparision} (see also Table~\ref{t:dents}) because the \cii\,emission of the Veil and the OMC dominate the \cii\,emission of Dents 1$-$4. 

\section{Summary}\label{Sect:Discussion}

    Using SOFIA \cii\,observations, we trace the influence of protostellar feedback by protostars on the Orion Veil. To that aim, we employ PV diagrams and maps of blue-shifted \cii\,emission ranging from $v_\mathrm{LSR}$ = $-$3 to $-$20~km~s$^{-1}$. A dent is defined as a shock-accelerated structure that expands outward on the Veil shell. We identify six dents in the Veil shell that are expanding towards us. Their sizes vary between 0.16 and 0.43~pc and they expand at velocities from 4 to 14~km~s$^{-1}$. \citet{Kavak2022} found that fossil outflows, generated by the Trapezium stars during their protostellar phase, influence the shape of the Veil shell as well. The momentum of the dents indicates newly forming stars with luminosities between 10$^3$ and 10$^4$~$L_\odot$, i.e., B-type stars. The dents are, therefore, a consequence of the collision of active, energetic jets/outflows expelled by massive protostars with the surrounding swept-up shell. The Veil shell is being driven mainly by the stellar wind of $\theta^1$~Ori~C, the most dominant star in the Trapezium cluster \citep{Abel2019, Pabst2019}. The Trapezium stars are on the main sequence and this wind is the result of radiation pressure acting on gas in the stellar photosphere. In contrast, the jets and outflows considered for the dents are driven by accretion onto a protostar. We conclude that, in addition to radiative feedback, both active and fossil outflow processes have a significant impact on the morphology of the Veil shell. 
    
    The total momentum in the dents are $\sim$1\% of the momentum carried by the Veil shell. \citet{Kavak2022} identified approximately twenty B-type stars in the direction of the Orion Nebula. Half of the twenty stars appear to be located within the Orion Nebula. We discovered six dents in all, and may missed four of them. These, we believe, could be located in the direction of the Huygens regions, which have a complex morphology. We also speculate that these protostellar jets and outflows may create channels and holes in the Veil that might allow the 10$^6$~K plasma to escape the Veil confinement. Any escaping hot plasma will entrain further Veil material and widen the dent aperture. Eventually, the escape of the hot plasma will relieve the pressure of the wind-blown bubble. At that point, the expansion of the Veil will enter a momentum conserving phase and eventually merge with the material in the Orion-Eridanus superbubble. Supernova explosions in the Orion Ia/Ib associations will sweep up this loose material and transport it to the walls of this superbubble \citep{Ochsendorf2015}. Further evidence is required to support this hypothesis.

    According to \citet{Vanderwerf2013, Abel2019}, the Veil shell is ionized by the Trapezium stars and has a multilayered structure along the line-of-sight. Because shocks in jets/outflows with high velocities in low-density slabs \citep[see also ][]{Lehmann2020} interact with the Veil, they may cause extra \cii\,emission on the Veil surface. This might be attributed to a variety of factors. To begin with, the Veil shell has a low, varying $N_\mathrm{H}$ toward the line-of-sight as it shows also a density gradient, which might suggest that there is insufficient material on the Veil surface to excite. In addition, the dents are positioned in front of the background OMC-1 cloud, which is likewise exposed to intense UV-radiation of Trapezium stars. This radiation may dominate additional \cii\,emission produced as a result of shock-cloud interaction.

    Without velocity-resolved \cii\,observations, it is challenging to unveil the dent-like structure on the Veil shell. Estimating the driving stars of the dents is also difficult as that star could have moved from the ejection point of its jets/outflows. Unlike CO-globules found in the Veil shell \citep{Goicoechea2020}, the dents do not show up in $^{12}$CO-PV diagrams. This shows that, like the Veil itself, their $N_\mathrm{H}$ is low because the dents are accelerated from the Veil.
        
    In the future, we plan to search for alternative tracers to follow the dents of the Veil and validate the presence of jets/outflows at their location. For this purpose, long-slit spectra of 1.644~$\mu$m \feii\,line, and \sii\,as can for example be observed with the ARCTIC instrument employed at Apache Point Observatory with high-resolution (>10,000) might aid in determining the dynamics of the dents. Finally, we conclude that velocity-resolved \cii\,observations of SOFIA observatory continue to be state-of-the-art for discovering feedback mechanisms in massive star forming regions.
    
    \section{Discussion}
    
    The Orion Nebula is the closest massive star-forming region, providing an excellent opportunity to study fundamental stellar feedback mechanisms. Recently SOFIA observations revealed that the Orion Veil shell is primarily driven by the stellar winds of $\theta^1$~Ori~C, an O6 type star in main-sequence-phase, and that the northern half of the Veil shell was damaged by its now-extinct outflows. While a single massive star that is $\theta^1$~Ori~C mainly dominates the most energetic mechanisms in Orion, other regions may include a varying number of massive stars as well as newly formed stars of different masses. Other massive star-forming regions are located further away from Orion, posing problems with spatial resolution and velocity components along the line-of-sight. The size of the dents varies by approximately 0.3~pc, and this type of structure may be resolved in Doppler space in a region five times the distance from the Orion Nebula. We will use velocity-resolved \cii\,line observation to search for dent-like structures in massive star forming regions to see if they are common in star-forming regions. In this context, the SOFIA FEEDBACK Legacy program \citep{Schneider2020} will provide \cii\,line observations of eleven massive star-forming regions at 158~$\mu$m.
    
\begin{acknowledgements}
    We thank Antoine Gusdorf for useful discussions on the shock models and Paul van der Werf for providing \hi\,21~cm observation of the Orion Veil. Studies of interstellar dust and gas at Leiden Observatory are supported by a Spinoza award from the Dutch Science agency, NWO. JRG thanks the Spanish MICINN for funding support under grant PID2019-106110GB-I00. This study was based on observations made with the NASA/DLR Stratospheric Observatory for Infrared Astronomy (SOFIA). SOFIA is jointly operated by the Universities Space Research Association Inc. (USRA), under NASA contract NAS2-97001, and the Deutsches SOFIA Institut (DSI), under DLR contract 50 OK 0901 to the University of Stuttgart. upGREAT is a development by the MPI für Radioastronomie and the KOSMA/Universität zu Köln, in cooperation with the DLR Institut für Optische Sensorsysteme. We acknowledge the work, during the C+ upGREAT square degree survey of Orion, of the USRA and NASA staff of the Armstrong Flight Research Center in Palmdale, the Ames Research Center in Mountain View (California), and the Deutsches SOFIA Institut. 
\end{acknowledgements}

%
%
\bibliographystyle{aa}
\bibliography{Kavaketal2022_VeilDents}

\begin{appendix}

\onecolumn

\section{Gaussian Fitting Results}

    \begin{table*}[h!]
    \caption{Fit results of multi-Gaussian fitting to \cii\,line profiles in Fig.~\ref{fig:dentspectrum}.}
    \label{t:fitresults}
    \centering
    \begin{tabular}{c c c c c c}
    \hline\hline
    &
    & $v_\mathrm{LSR}$
    & $\int$~$T_\mathrm{mb}\delta$V
    & $\Delta$V
    & $T_\mathrm{mb}$
    \\
    Component
    & Position
    & [km~s$^{-1}$]
    & [K~km~s$^{-1}$]
    & [km~s$^{-1}$]
    & [K] 
    \\
    \hline\hline
        Dent & 1  & $-$17.5~$\pm$~0.81 & 5.27~$\pm$~1.51    &   5.77~$\pm$~1.92  &    0.85~$\pm$~0.25  \\
             & 2  & $-$9.59~$\pm$~3.34 & 9.91~$\pm$~4.23    &   15.8~$\pm$~6.46  &    0.58~$\pm$~0.09  \\
             & 3  & $-$7.09~$\pm$~1.55 & 11.6~$\pm$~3.18    &   11.6~$\pm$~2.54  &    0.94~$\pm$~0.09  \\
             & 4  & $-$12.8~$\pm$~1.63 & 6.27~$\pm$~1.50    &   14.3~$\pm$~4.01  &    0.41~$\pm$~0.06  \\
             & 5  & $-$7.62~$\pm$~0.19 & 22.8~$\pm$~1.25    &   7.12~$\pm$~0.47  &    3.00~$\pm$~0.16  \\
             & 6  & $-$9.23~$\pm$~0.29 & 6.23~$\pm$~3.61    &   3.86~$\pm$~0.84  &    1.51~$\pm$~0.58  \\
    \hline
        Veil & 1  & $+$1.02~$\pm$~0.27 & 8.02~$\pm$~1.36    &   3.50~$\pm$~0.68  &    2.15~$\pm$~0.31  \\
             & 2  & $+$0.01~$\pm$~0.10 & 34.7~$\pm$~3.52    &   6.16~$\pm$~0.33  &    5.29~$\pm$~0.32  \\
             & 3  & $+$0.90~$\pm$~0.21 & 18.6~$\pm$~3.07    &   6.67~$\pm$~0.49  &    2.62~$\pm$~0.28  \\
             & 4  & $+$0.63~$\pm$~0.12 & 30.9~$\pm$~1.37    &   7.76~$\pm$~0.31  &    3.74~$\pm$~0.09  \\
             & 5  & $+$0.92~$\pm$~0.48 & 16.8~$\pm$~2.06    &   9.96~$\pm$~1.46  &    1.59~$\pm$~0.10  \\
             & 6  & $-$4.91~$\pm$~1.98 & 6.74~$\pm$~3.82    &   7.23~$\pm$~2.91  &    0.87~$\pm$~0.18  \\
    \hline
        OMC  & 1  & $+$9.21~$\pm$~0.02 & 146.7~$\pm$~1.3    &   4.18~$\pm$~0.04  &    32.9~$\pm$~0.29  \\
             & 2  & $+$10.2~$\pm$~0.06 & 29.43~$\pm$~0.8    &   4.42~$\pm$~0.01  &    6.24~$\pm$~0.16  \\
             & 3  & $+$9.76~$\pm$~0.01 & 106.5$\pm$~0.44    &   3.70~$\pm$~0.02  &    27.0~$\pm$~0.10  \\
             & 4  & $+$9.54~$\pm$~0.02 & 51.0~$\pm$~0.55    &   3.62~$\pm$~0.04  &    13.2~$\pm$~0.12  \\
             & 5  & $+$9.33~$\pm$~0.02 & 23.0~$\pm$~0.58    &   1.87~$\pm$~0.04  &    11.5~$\pm$~0.22  \\
             & 6  & $+$8.81~$\pm$~0.61 & 13.6~$\pm$~0.46    &   3.71~$\pm$~0.14  &    3.45~$\pm$~0.11  \\
    \hline
    \end{tabular}
\end{table*}

\section{Massive Stars and Geometry}

        \begin{figure}[!ht]
            \centering
            \includegraphics[width = 0.5 \columnwidth]{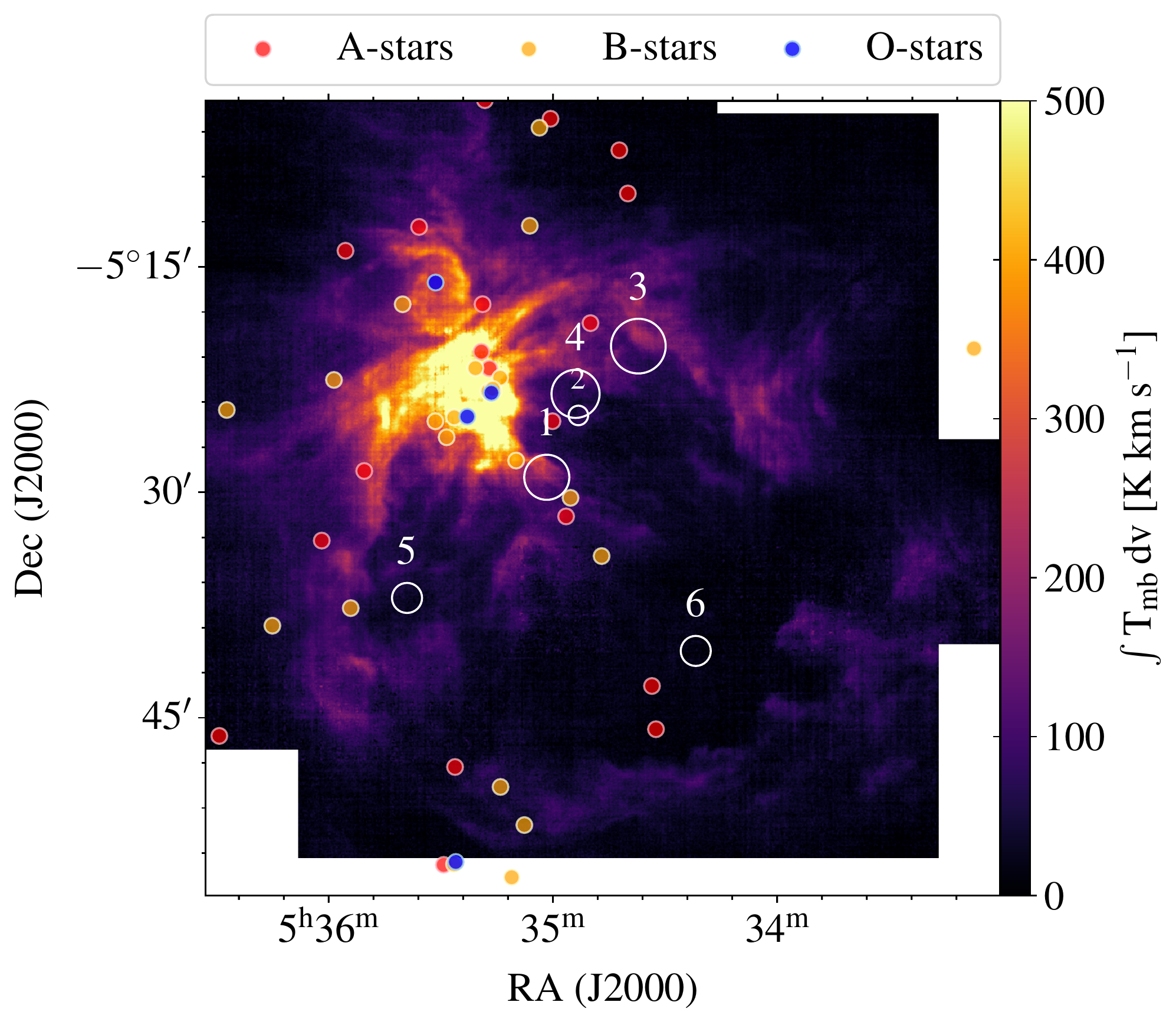}
            \includegraphics[width = 0.4 \columnwidth]{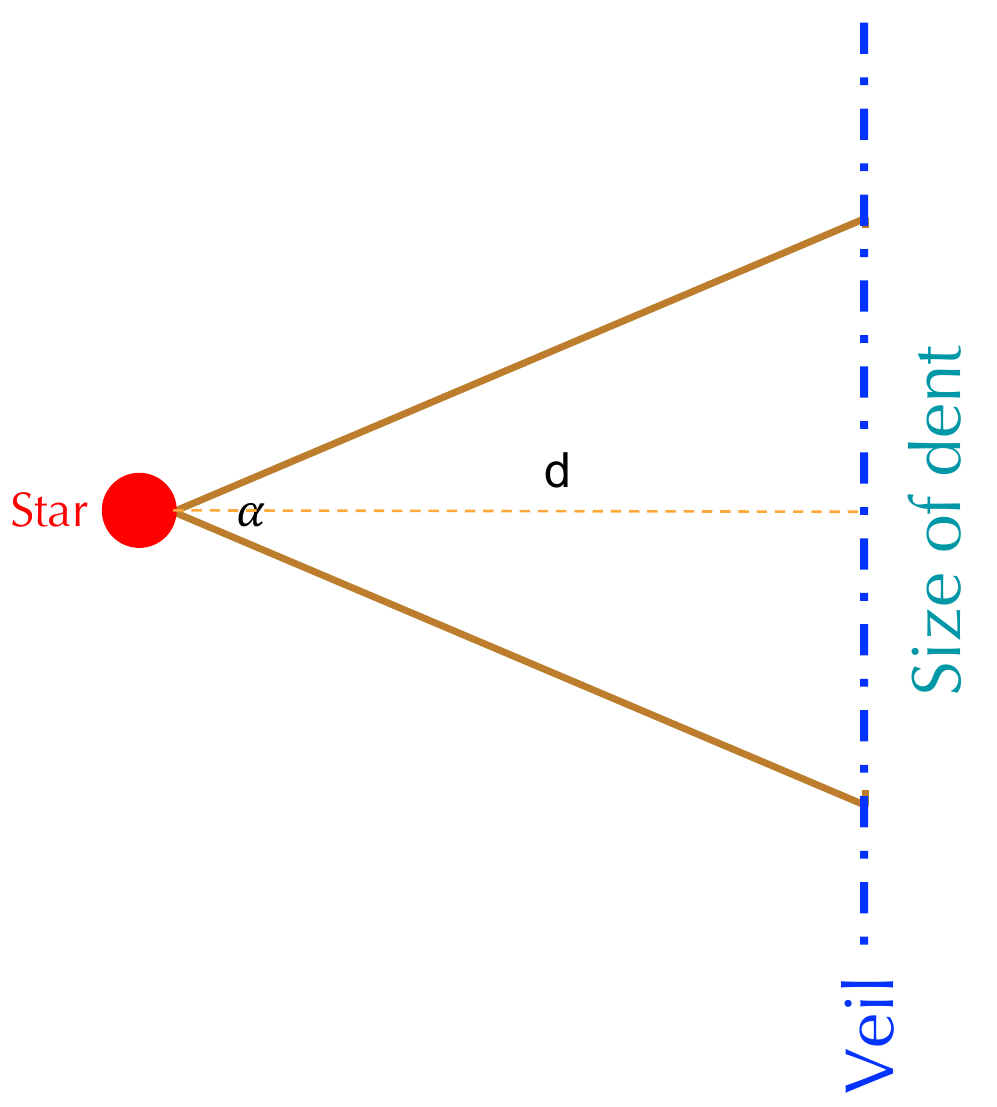}
            \caption{\textit{Left}: SOFIA \cii\,map of Orion with O$-$, B$-$, and A$-$stars found in SIMBAD. The blue$-$, orange$-$, and red$-$filled circles are O$-$, B$-$, and A$-$stars, respectively. White open circles indicate the dents identified in this work. \textit{Right}: Geometry we used to calculate the collimation factor and opening angle ($\alpha$).}
            \label{fig:OBAstarsDents}
        \end{figure}

\section{PV diagram of the dents}\label{sect:dentpvdiagrams}

    This section contains a series of PV diagrams that cover the dents studied in this work. The length of all PV diagrams is 60$\arcmin$. Dents are indicated by a colored arrow.

\begin{figure*}
    \centering
    \includegraphics[width = \columnwidth]{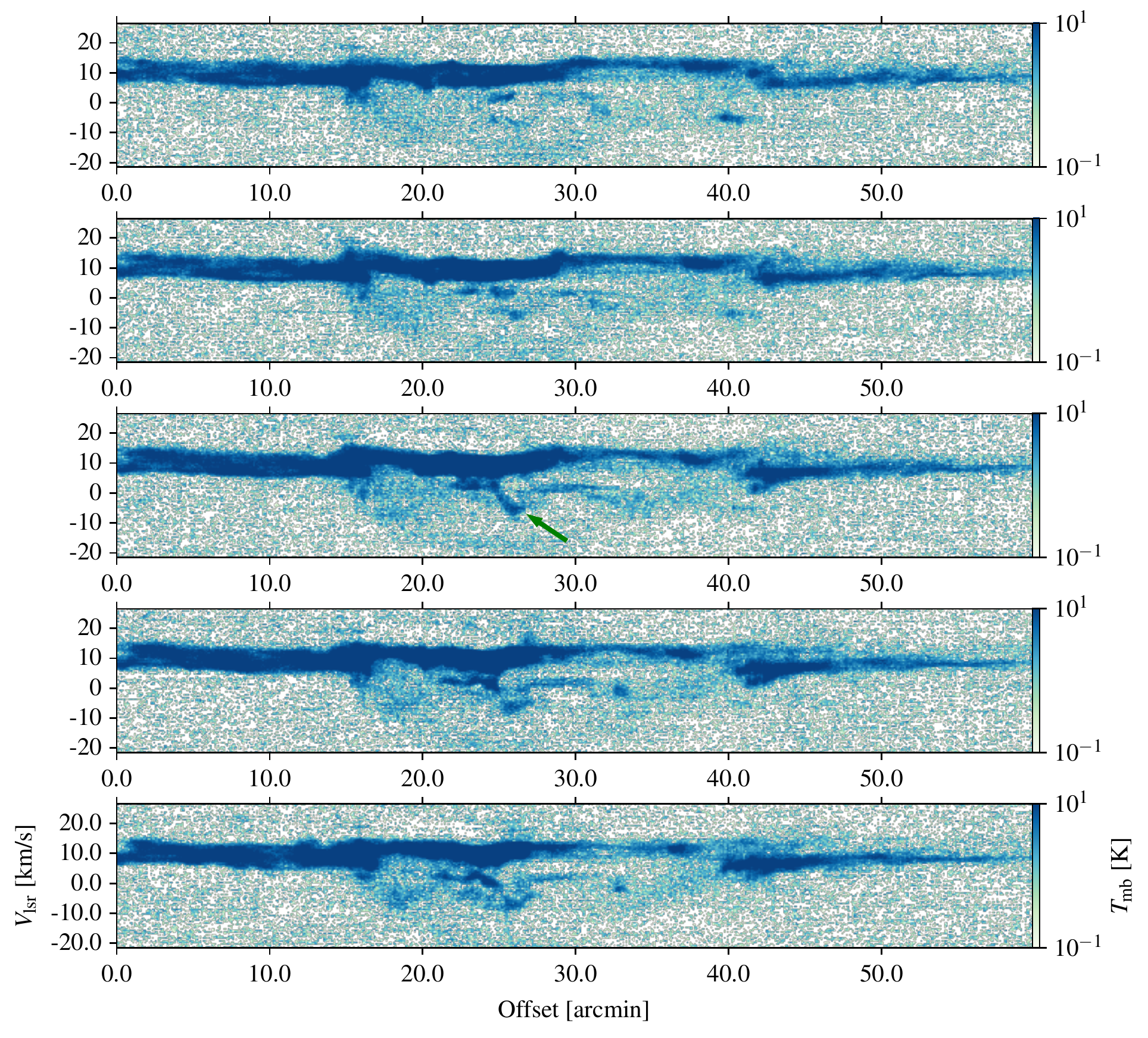}
    \caption{Five consecutive PV diagrams showing the changes of the dent 1. The PV diagrams cover the Veil shell from east to west, spanning 60$\arcmin$ in length and 30$\arcsec$ in width. The declination of the PV diagram changes from top to bottom panel. The dent at 25$\arcmin$ is indicated with a green arrow.}
    \label{fig:dent1}
\end{figure*}

    \begin{figure*}[!ht]
    \centering
            \includegraphics[width = \columnwidth]{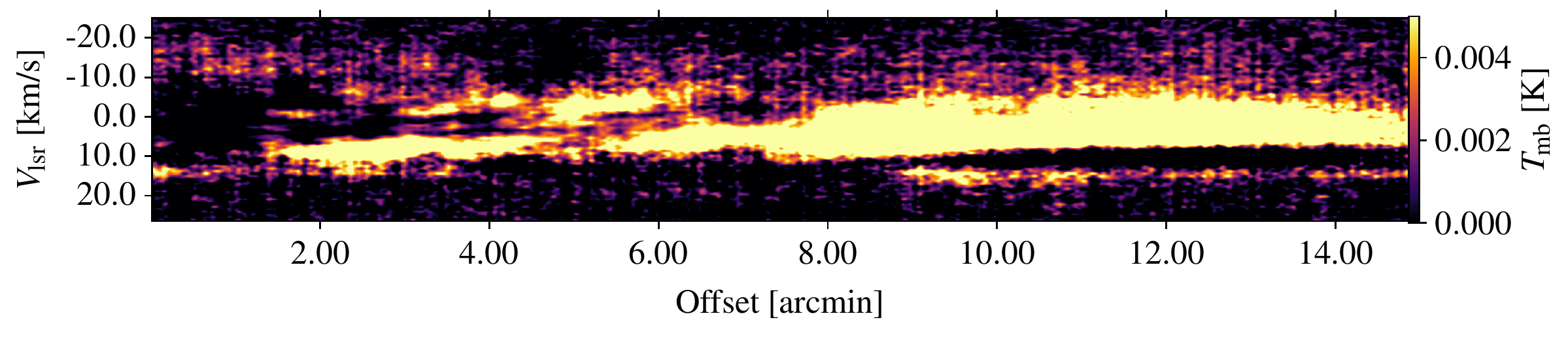}
            \caption{\hi\,21 cm PV diagram of dent 1. On Fig.~\ref{fig:dent1}, the dent indicated by a green arrow is centered on the center of the PV diagram, i.e., at 7.5$\arcmin$ in the x-axis. \hi\,observation exists between the offset 18$\arcmin$ and 33$\arcmin$ in the \cii\,pv-diagram in Fig.~\ref{fig:dent1}.}
            \label{fig:dent1_hi}
    \end{figure*}

\begin{figure*}
    \centering
    \includegraphics[width = \columnwidth]{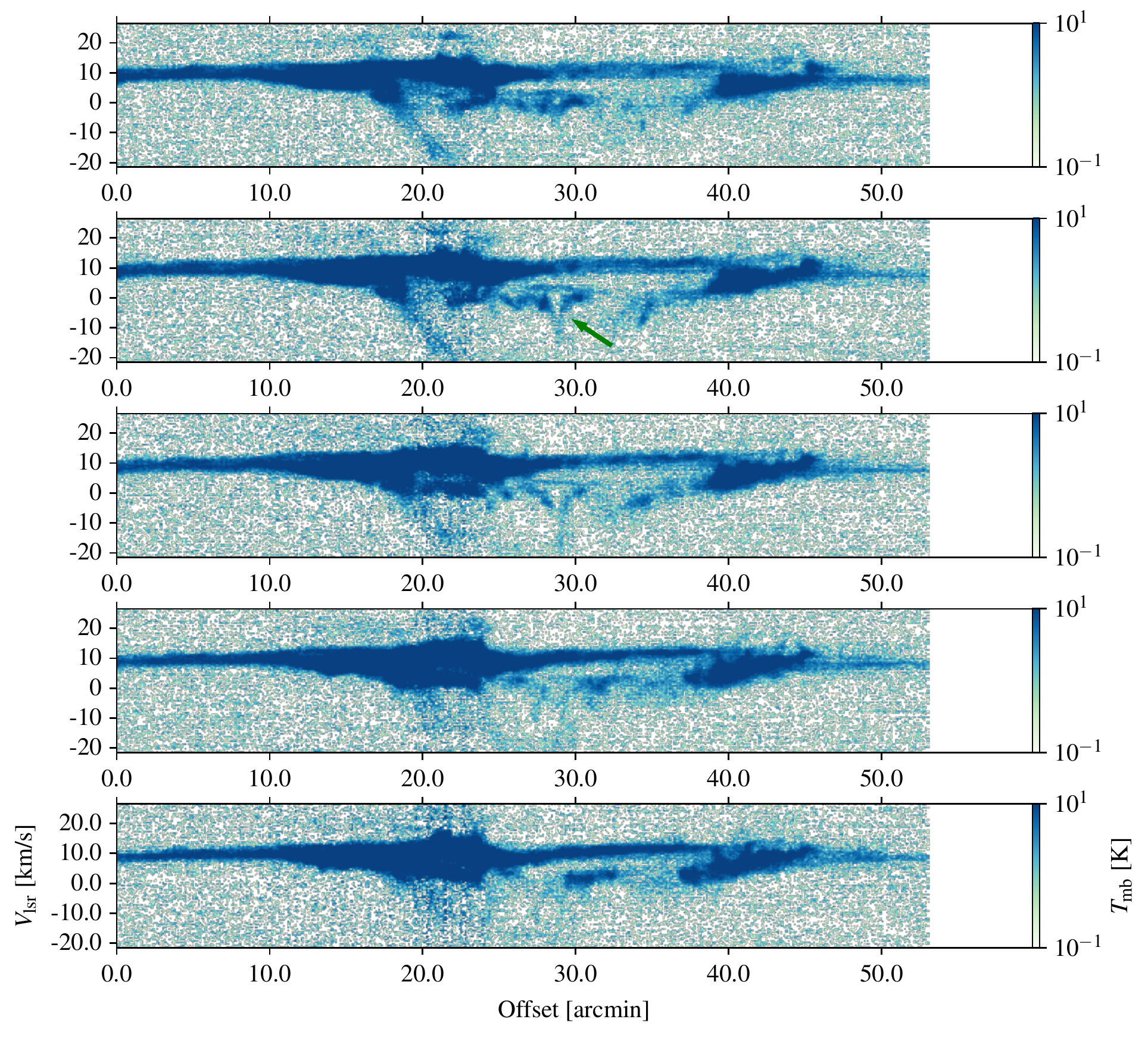}
    \caption{Five consecutive PV diagrams showing the changes of the dent 2. The PV diagrams cover the Veil shell from east to west, spanning 60$\arcmin$ in length and 30$\arcsec$ in width. The declination of the PV diagram changes from top to bottom panel. The dent at 27$\arcmin$ is indicated with a green arrow.}
    \label{fig:dent2}
\end{figure*}

    \begin{figure*}[!ht]
        \centering
            \includegraphics[width = \columnwidth]{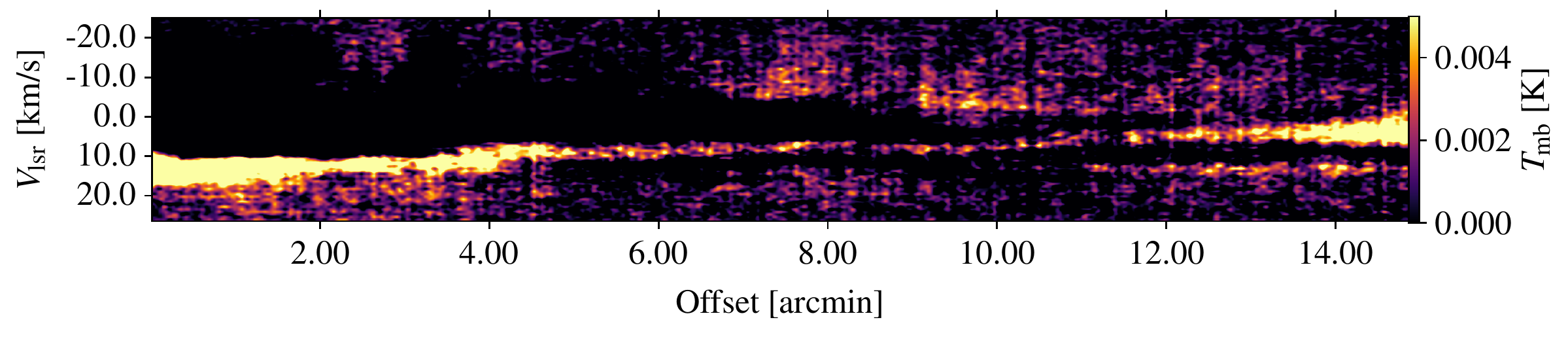}
            \caption{\hi\,21 cm PV diagram of dent 2. On Fig.~\ref{fig:dent2}, the dent indicated by a green arrow is centered on the center of the PV diagram, i.e., at 7.5$\arcmin$ in the x-axis. \hi\,observation exists between the offset 20$\arcmin$ and 35$\arcmin$ in the \cii\,PV-diagram in Fig.~\ref{fig:dent2}.}
            \label{fig:dent2_hi}
    \end{figure*}

\begin{figure*}
    \centering
    \includegraphics[width = \columnwidth]{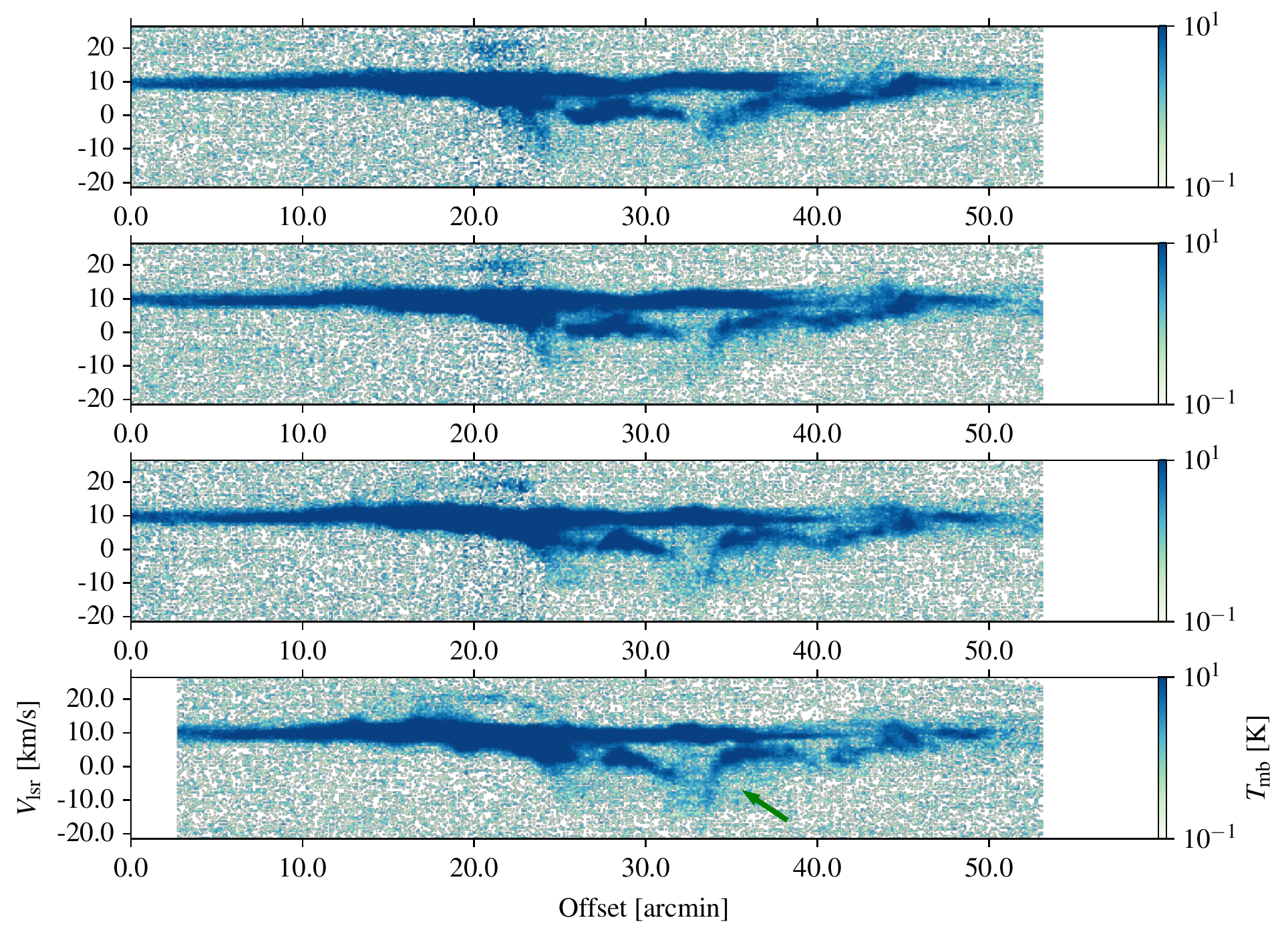}
    \caption{Four consecutive PV diagrams showing the changes of the dent 3. The PV diagrams cover the Veil shell from east to west, spanning 60$\arcmin$ in length and 30$\arcsec$ in width. The declination of the PV diagram changes from top to bottom panel. The dent at 33$\arcmin$ is indicated with a green arrow.}
    \label{fig:dent3}
\end{figure*}

    \begin{figure*}[!ht]
            \includegraphics[width = \columnwidth]{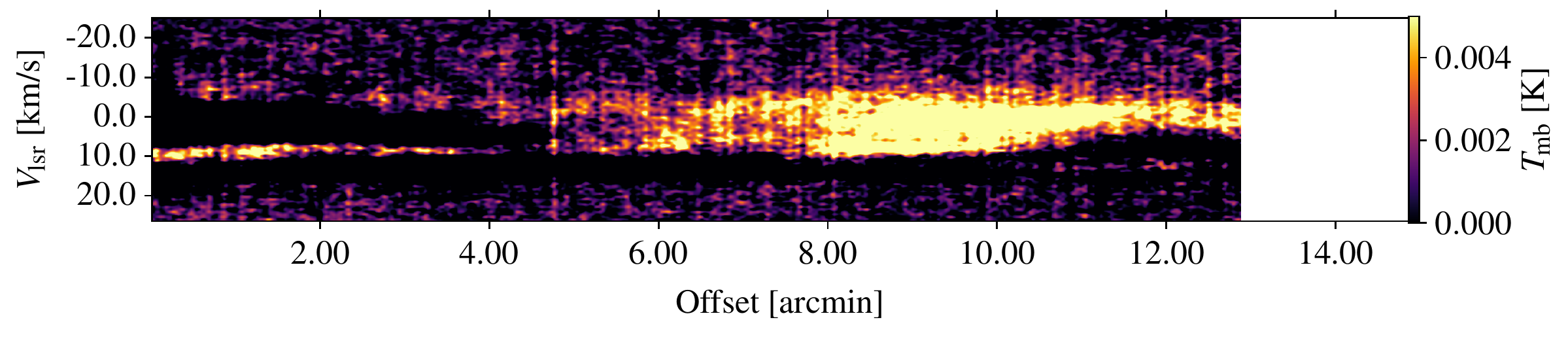}
            \caption{\hi\,21 cm PV diagram of dent 3. On Fig.~\ref{fig:dent3}, the dent indicated by a green arrow is centered on the center of the PV diagram, i.e., at 7.5$\arcmin$ in the x-axis. \hi\,observation exists between the offset 25$\arcmin$ and 40$\arcmin$ in the \cii\,PV-diagram in Fig.~\ref{fig:dent3}.}
            \label{fig:dent3_hi}
    \end{figure*}

\begin{figure*}
    \centering
    \includegraphics[width = \columnwidth]{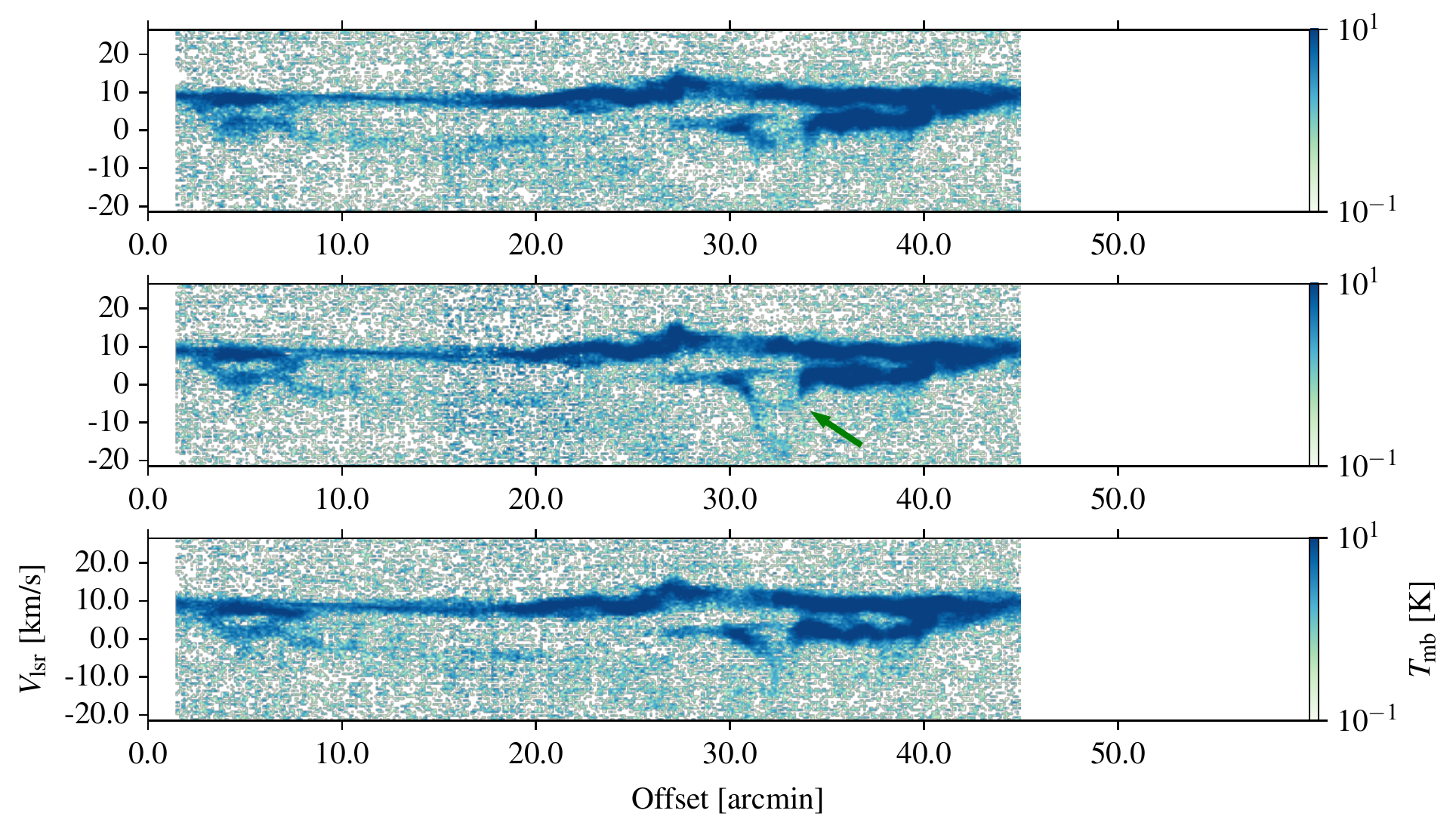}
    \caption{Three consecutive PV diagrams showing the changes of the dent 4. The PV diagrams cover the Veil shell from east to west, spanning 60$\arcmin$ in length and 30$\arcsec$ in width. The declination of the PV diagram changes from top to bottom panel. The dent at 34$\arcmin$ is indicated with a green arrow.}
    \label{fig:dent4}
\end{figure*}

    \begin{figure*}[!ht]
            \includegraphics[width = \columnwidth]{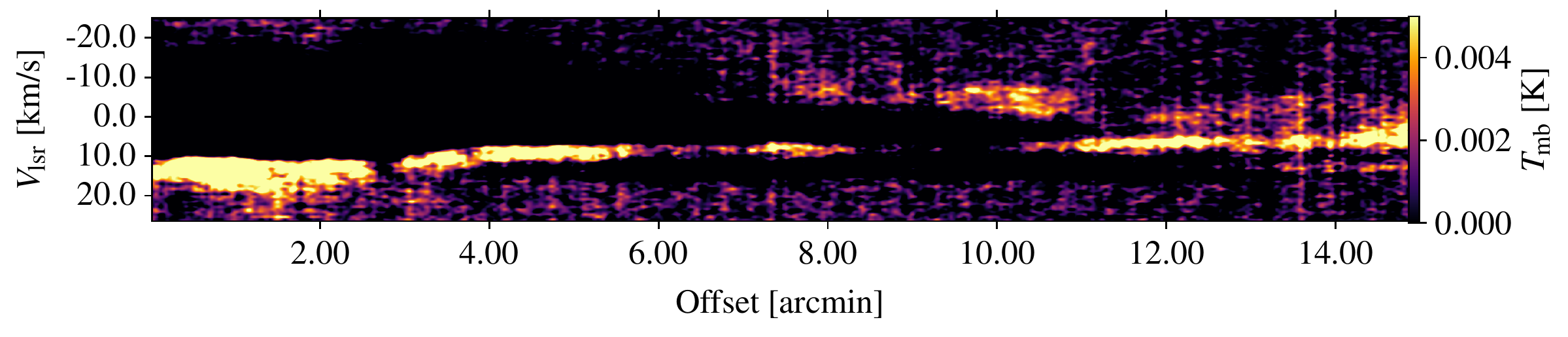}
            \caption{\hi\,21 cm PV diagram of dent 4. On Fig.~\ref{fig:dent4}, the dent indicated by a green arrow is centered on the center of the PV diagram, i.e., at 7.5$\arcmin$ in the x-axis. \hi\,observation exists between the offset 25$\arcmin$ and 40$\arcmin$ in the \cii\,PV-diagram in Fig.~\ref{fig:dent4}.}
            \label{fig:dent4_hi}
    \end{figure*}

\begin{figure*}
    \centering
    \includegraphics[width = \columnwidth]{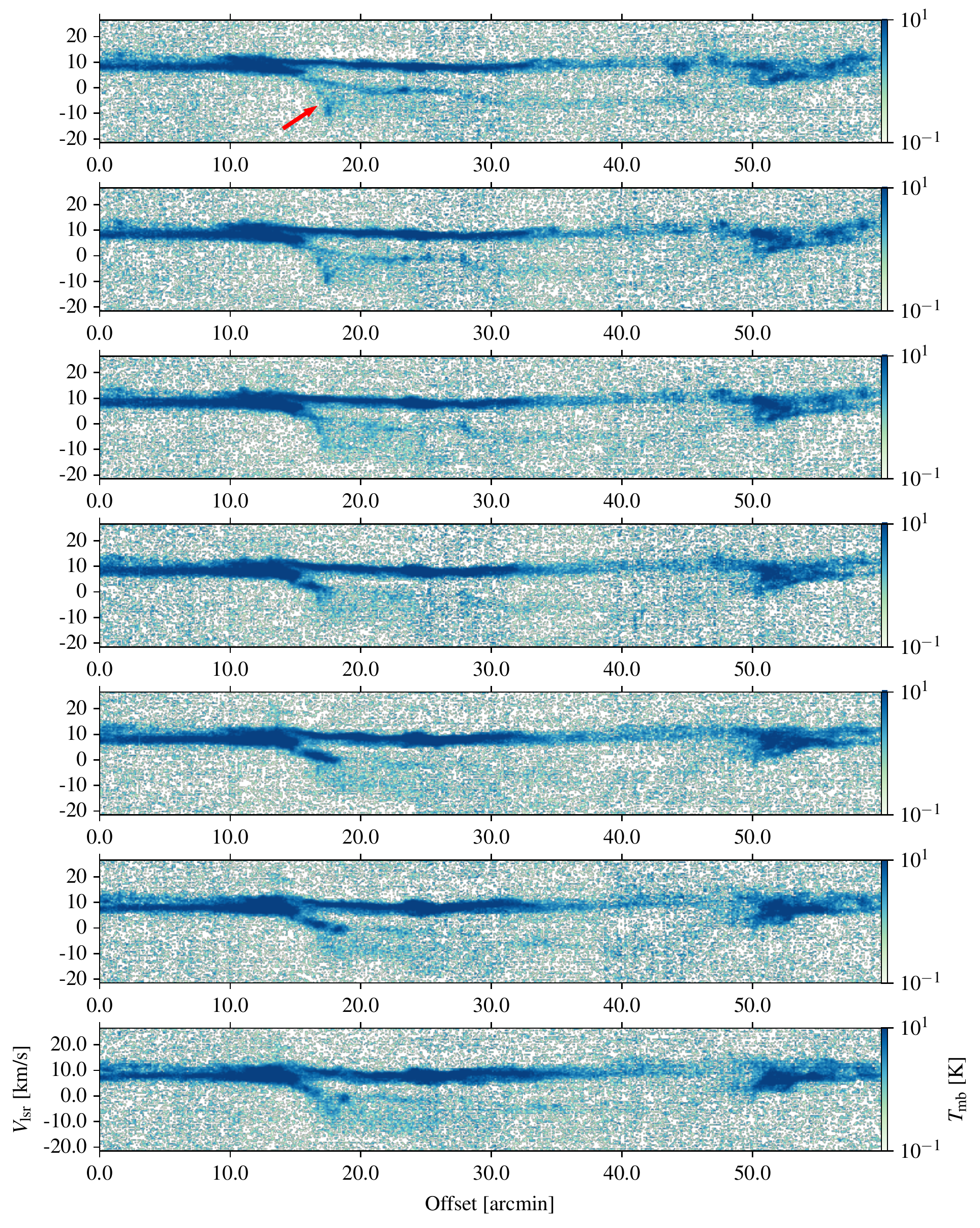}
    \caption{Seven consecutive PV diagrams showing the changes of the dent 5. The PV diagrams cover the Veil shell from east to west, spanning 60$\arcmin$ in length and 30$\arcsec$ in width. The declination of the PV diagram changes from top to bottom panel. The dent at 18$\arcmin$ is indicated with a red arrow.}
    \label{fig:Dent5} 
\end{figure*}

\begin{figure*}
    \centering
    \includegraphics[width = \columnwidth]{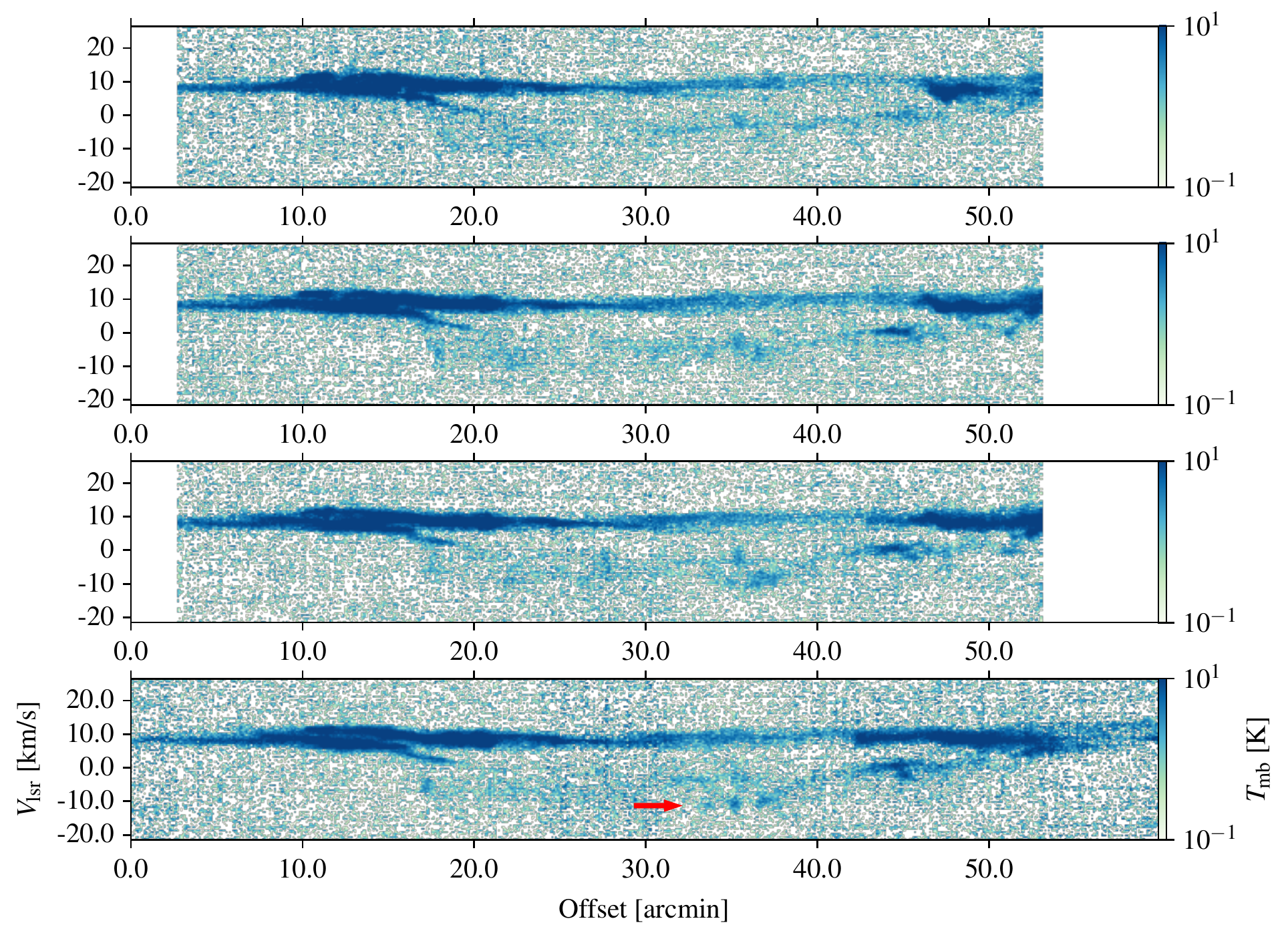}
    \caption{Four consecutive PV diagrams showing the changes of the dent 6. The PV diagrams cover the Veil shell from east to west, spanning 60$\arcmin$ in length and 30$\arcsec$ in width. The declination of the PV diagram changes from top to bottom panel. The dent at 33$\arcmin$ is indicated with a red arrow.}
    \label{fig:Dent6}
\end{figure*}

\begin{figure*}
    \centering
    \includegraphics[width = 0.411 \columnwidth]{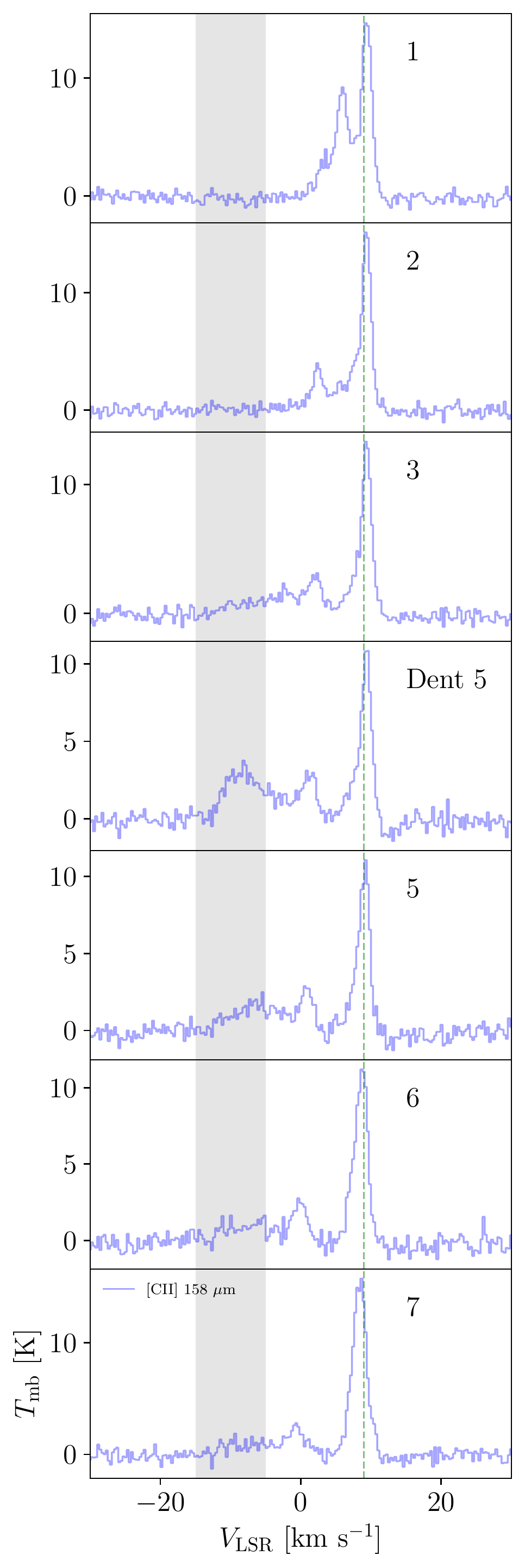}
    \includegraphics[width = 0.4 \columnwidth]{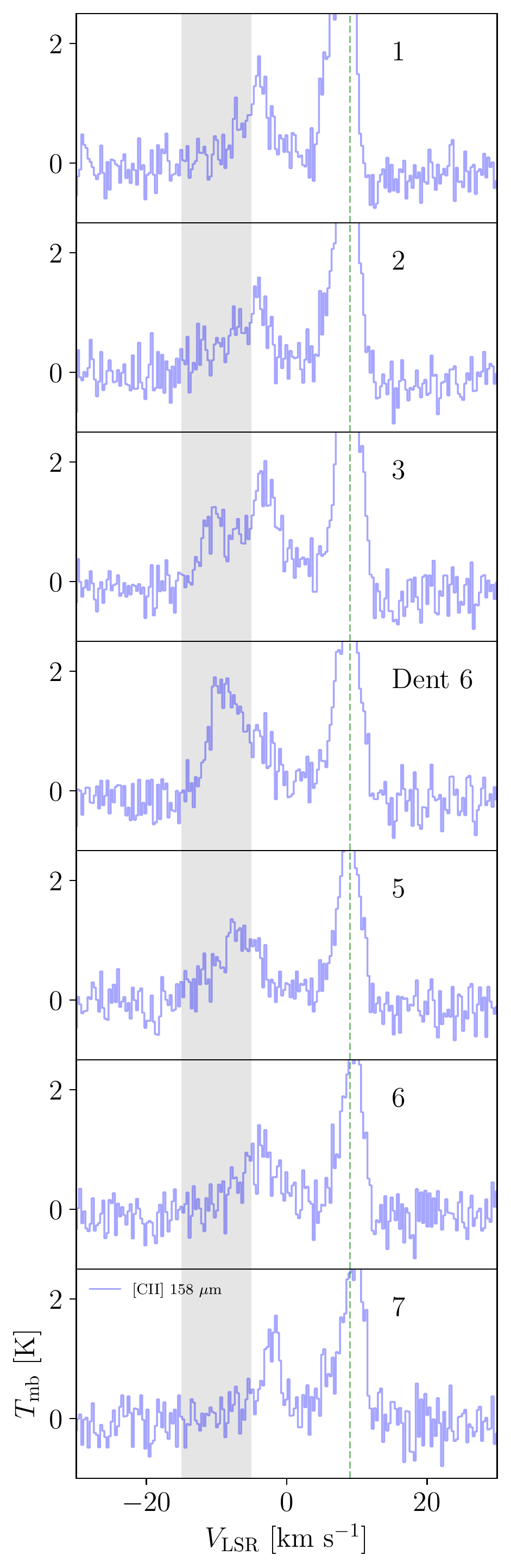}
    \caption{Horizontally consecutive \cii\,spectra extracted over dents 5 and 6, demonstrating the change of the line profile.}
    \label{fig:spectrum5_6}
\end{figure*}

\end{appendix}

\end{document}